\documentclass{ieeeaccess}
\usepackage[ruled]{algorithm2e}
\usepackage{amsfonts}
\usepackage{amssymb}
\usepackage{eurosym}
\usepackage{authblk}
\usepackage{cite}
\usepackage{graphicx}
\usepackage{amsmath}
\usepackage[labelformat=simple]{subcaption}
\usepackage{float}
\usepackage[T1]{fontenc}
\usepackage{slashbox}
\usepackage{amsthm}
\usepackage{mathrsfs}
\usepackage{lipsum}
\usepackage{mathtools}
\usepackage{multicol}
\usepackage{textcomp}
\usepackage{setspace}
\usepackage{listings}
\usepackage{color}
\usepackage{hyperref}
\usepackage{cleveref}

\definecolor{mygreen}{RGB}{28,172,0}
\definecolor{mylilas}{RGB}{170,55,241}
\def\BibTeX{{\rm B\kern-.05em{\sc i\kern-.025em b}\kern-.08em
    T\kern-.1667em\lower.7ex\hbox{E}\kern-.125emX}}

\newtheorem{Remark}{Remark}
\newcounter{mytempeqncnt}

\DeclarePairedDelimiter\floor{\lfloor}{\rfloor}

\begin{document}
\lstset{language=Matlab,
           basicstyle=\fontsize{8}{8} \ttfamily,
           keywordstyle=\color{blue} \fontsize{8}{8}  \ttfamily,
           stringstyle=\color{mylilas} \fontsize{8}{8} \ttfamily,
           commentstyle=\color{mygreen} \fontsize{8}{8}  \ttfamily,
          breaklines=true,
        }

\history{Date of publication xxxx 00, 0000, date of current version xxxx 00, 0000.}
\doi{00.0000/ACCESS.2019.DOI}
\title{An Improved Accurate Solver for the Time-Dependent RTE in Underwater
Optical Wireless Communications}
\author{\uppercase{Elmehdi Illi}\authorrefmark{1}, \IEEEmembership{Student Member, IEEE},
\uppercase{Faissal El Bouanani}\authorrefmark{1}, \IEEEmembership{Member, IEEE}, \uppercase{Ki-Hong Park}\authorrefmark{2}, \IEEEmembership{Member, IEEE}, \uppercase{Fouad Ayoub}\authorrefmark{3}, \IEEEmembership{Member, IEEE}, and \uppercase{Mohamed-Slim Alouini}\authorrefmark{2},
\IEEEmembership{Fellow, IEEE}}

\address[1]{ENSIAS, Mohammed V University in Rabat, Morocco (e-mails: \{elmehdi.illi, f.elbouanani\}@um5s.net.ma)}
\address[2]{Computer, Electrical, and Mathematical Sciences and Engineering (CEMSE) Division, King Abdullah University of Science and Technology (KAUST), Thuwal 23955-6900, Saudi Arabia
(e-mails:\{kihong.park, slim.alouini\}@kaust.edu.sa).}
\address[3]{CRMEF, Kenitra, Morocco (e-mail: ayoub@crmefk.ma )}

\markboth
{E. Illi \headeretal: An Improved Accurate Solver for the Time-Dependent RTE in Underwater
Optical Wireless Communications}
{E. Illi \headeretal: An Improved Accurate Solver for the Time-Dependent RTE in Underwater
Optical Wireless Communications}

\corresp{Corresponding author: Faissal El Bouanani (e-mail: f.elbouanani@um5s.net.ma).}

\begin{abstract}
In this paper, an improved numerical solver to evaluate the time-dependent
radiative transfer equation (RTE) for underwater optical wireless
communications (UOWC) is investigated. The RTE evaluates the optical
path-loss of light wave in an underwater channel in terms of the inherent
optical properties related to the environments, namely the absorption and
scattering coefficients as well as the phase scattering function (PSF). The
proposed numerical algorithm was improved based on the ones proposed in \cite%
{klose, gao, alouini, commnet}, by modifying the finite difference scheme
proposed in \cite{klose} as well as an enhancement of the quadrature method
proposed in \cite{gao} by involving a more accurate 7-points quadrature
scheme in order to calculate the quadrature weight coefficients
corresponding to the integral term of the RTE. Furthermore, the scattering
angular discretization algorithm used in \cite{alouini} and \cite{commnet}
was modified, based on which the receiver's field of view discretization was
adapted correspondingly. Interestingly, the RTE solver has been applied to
three volume scattering functions, namely: the single-term HG phase
function, the two-term HG phase function \cite{kattawar}, and the
Fournier-Forand phase function \cite{fournierorg}, over Harbor-I and
Harbor-II water types. Based on the normalized received power evaluated
through the proposed algorithm, the bit error rate performance of the UOWC
system is investigated in terms of system and channel parameters. The
enhanced algorithm gives a tightly close performance to its Monte Carlo
counterpart improved based on the simulations provided in \cite{cox}, by
adjusting the numerical cumulative distribution function computation method
as well as optimizing the number of scattering angles. Matlab codes for the
proposed RTE solver are presented in \cite{github}.
\end{abstract}

\titlepgskip=-15pt
\maketitle

\section{ Introduction}

With the prospering of the wireless communication industry over the last
decades, human exploration in the underwater environment increased
significantly. More recently, a notable increase in research activities in
the marine medium has been witnessed, which enabled the deployment of ocean
exploration and communication systems \cite{laura}, \cite{survey}. By this
period, several emerging underwater applications have attracted many
interests, such as climate recording, ecological monitoring, oil production
control, and military surveillance \cite{access}. With this permanent
emphasis on researches in the marine medium, underwater wireless sensors
network concept came on for enabling the concretization of several critical
commercial and military applications and services \cite{kaushal}. In
particular, wireless communication nodes such as wireless sensors, floating
buoys, and submarines, require reliable links with higher data rates in
order to fulfill communication requirements and exchange a relatively huge
amount of data \cite{surveyal}.

Nowadays, underwater wireless communication systems (UWCS) are implemented
using acoustic, radio-frequency (RF), or optical wireless communication.
Conventional RF communication technology based on carrier-modulated signals
are seriously affected by extremely severe attenuation at high frequencies
due to the high concentration of salt, which is a conductive material \cite%
{survey}, \cite{arnon}, e.g., the attenuation in ocean water can reach 169
dB/m for 2.4 GHz band \cite{surveyal}. Moreover, RF-based UWC devices demand
heavy and energy-consuming antennas, making RF\ technology an impractical
candidate for UWCS \cite{survey}. On the contrary, acoustic technology is by
far the most deployed technology nowadays in the UWCS \cite{milica}, as it
provides a much longer transmission link range compared to its RF
counterpart (i.e., up to 20 km) \cite{arnon2}. Nevertheless, acoustic
communication is affected by some technical limitations; due to the low
sound wave celerity in the water (i.e., 1500 m/s at the pure water), the
acoustic link suffers from serious communication delays called latency \cite%
{illi}. Additionally, as the frequency range associated with acoustic
communication is between tens and hundreds of Kilohertz, the available
communication data rate is relatively low (e.g., in the order of Kbps).
Also, similarly to their RF counterparts, acoustic transceivers are heavy,
costly, and energy-consuming \cite{survey}.

Optical wireless communication (OWC) is an emerging technology that received
considerable attention lastly, as a promising key-enabling technology for
high-speed terrestrial and underwater communications \cite{access}. It
consists of transmitting the information signals in the form of light
conical beams using LED\ or laser devices through either the free space;
i.e. visible light communication (VLC), free space optics (FSO), or the
underwater medium (UOWC) \cite{arnon}, \cite{fso1}. Due to its great
potential for providing a tremendous amount of bandwidth, high security as
well as immunity to interference, OWC\ is the most advocated solution in
providing a low-latency communication link with data rates of tens of Gbps
over moderate distances \cite{khalighi}.

Generally, light propagation in the marine medium is corrupted by three main
phenomena: Stochastic phenomena, namely (i) turbulence-induced fading due to
sea movement as well as temperature and pressure inhomogeneities \cite%
{access}, (ii) pointing errors due to transceiver motion \cite{survey}. On
the other hand, (iii) path-loss is a deterministic phenomenon affecting
light propagation, caused mainly by photons absorption and scattering,
representing the two major inherent optical properties (IOP) that quantify
light power loss \cite{mobley}. Absorption is the process where the photons
lose their energy by conversion into another form such as chemical or heat,
while scattering indicates the photons direction change due to the light
interaction with the medium particles and molecules \cite{mobley}. That is,
the greater the scattering and absorption coefficients, the severer the
power loss in the medium.

Several approaches in the literature have been proposed to analyze and
predict the total light power path-loss in the marine medium. Beer-Lambert's
law is a deterministic approach, and it is the simplest model applied to
evaluate the optical loss \cite{survey}. Indeed, it considers an exponential
decay of the received light intensity as a function of the propagation
distance, attenuation coefficient, defined as the sum of absorption and
scattering coefficients, as well as source intensity. However, its main
drawback lies in assuming that the scattered photons are completely lost,
while in fact, some of them can still be captured at the receiver after
multiple scattering, and therefore the received power is underestimated \cite%
{alouini}. On the other hand, Monte Carlo simulation method is among popular
numerical approaches to evaluate the optical path loss in underwater medium.
It is a probabilistic method that emulates underwater light transmission
loss by transmitting and tracking the propagation of a huge number of
simulated photons \cite{cox}. Its main benefits lie in its easy
implementation in computation platforms, as well as its acceptable accuracy
often. However, its main limitation lies in the errors related to the random
values generators as well as its long running time \cite{survey}.

During the recent past years, the use of radiative transfer equation (RTE)\
has attracted significant attention in the fields of optics for biomedical
imaging \cite{ripoll}. In particular, it is considered as a deterministic
solution for describing light propagation in multiple absorbing and
scattering medium (e.g., fluids, underwater environment), in terms of the
medium IOP, such as absorption and scattering coefficients as well as the
phase scattering function (PSF). Interestingly, this latter defines the
scattering power distribution over the various directions in the propagation
medium. In this regard, the single-term Henyey-Greenstein (STHG) function
has been widely adopted as an analytical model for highly peaked forward
scattering environments \cite{alouini}. Nevertheless, due to the inaccurate
fitting of the STHG phase function with scattering measurements for most of
the realistic marine environments, the two-terms Henyey-Greenstein (TTHG)
and the Fournier-Forand (FF) phase functions have been advocated as
analytical models for the underwater PSF\ modeling.

Even though the RTE is already more than a century old, very few works along
this period involved this equation for evaluating the light power loss in
various scattering mediums, since it is enough complicated to solve the
integro-differential RTE analytically. Actually, various numerical
approaches have been used for this purpose, namely \cite{klose}, \cite{gao},
\cite{alouini}, and \cite{commnet}. In \cite{klose}, a numerical approach to
solve the time-dependent (TD) RTE using the finite difference equation and
the discrete ordinate method (DOM) was proposed. This latter consists of
discretizing the angular and spatial coordinates uniformly into finite
equally spaced points.\ Interestingly, trapeze quadrature method was used to
solve the RTE integral term. The authors in \cite{gao} deployed another
numerical approach for solving the steady-state time-independent (TI) RTE,
based also on the DOM and the upwind finite difference scheme for the
partial derivatives, as well as using the 3-points Simpson's quadrature
method to solve the integral term. In \cite{alouini}, an improvement was
made related to the numerical proposal in \cite{gao}, where an optimal
non-uniform angular discretization through the Lloyd-Max algorithm \cite%
{lloyd} is proposed. Furthermore, the Gauss-Seidel iterative method was
involved to solve the fully discretized system of linear equations. Finally,
the proposed solver in \cite{alouini} was improved in \cite{commnet} by
involving a two-neighbours derivative for space coordinates in addition to
involving time derivative, as well as incorporating the 5-points quadrature
scheme alongside with the 3-points one. It is noteworthy that the majority
of the previous related works dealt with the STHG\ as a PSF.

In this paper, an enhancement of the numerical TD-RTE solvers developed in
\cite{alouini} and \cite{commnet} is investigated. Distinctly to these
latter where the 3 and 5-points quadrature schemes were applied to
neighboring scattering angles, which is inaccurate, we involve in this paper
the 7-points quadrature scheme, which we apply to infinitesimally small
subintervals. Additionally, the scattering angles discretizing algorithm
used in the two abovementioned works was modified. The main contributions of
this paper are highlighted as follows:

\begin{itemize}
\item The 3 and 5-points quadrature methods used in \cite{alouini} and \cite%
{commnet}, respectively, are adjusted by involving the 7-point rule given by
the Newton-Cotes formula. The quadrature method aims at determining the
weight coefficients in order to compute the integral term of the TD-RTE.
Distinctly from \cite{alouini} and \cite{commnet}, where the above-mentioned
quadrature schemes were applied to neighboring scattering angles, which is
an inaccurate approach since the step between two angles is relatively
great, we propose to apply this interpolation method to the discretized
infinitesimally small subintervals within the interval between two
successive scattering angles.

\item Distinctly from \cite{alouini} and \cite{commnet}, the mean squared
error (MSE) based algorithm, used for the scattering angles discretization,
is modified. In particular, the updated version in this paper relaxes the
symmetric scattering distribution considered previously. Furthermore, the
receiver's field of view (FOV) has been discretized in a similar manner to
the scattering angles.

\item As performed in \cite{commnet}, a more accurate finite upwind
difference scheme, incorporating two neighbor points, is involved.

\item In addition to applying the TD-RTE\ solver to the classical STHG
function used in \cite{alouini} and \cite{commnet}, the TTHG and FF phase
functions are also adopted as PSFs in this work.

\item Monte Carlo simulation provided in \cite{cox} was updated, by
modifying the numerical cumulative distribution function (CDF)\ computation
as well as optimizing the number of generated scattering angles for each
subinterval of two successive distances.

\item The bit error rate (BER) performance is analyzed, based on the
evaluated received power, as a function of propagation time and distance as
well as the system and channel parameters.

\item The numerical RTE\ results are compared with their Monte Carlo (MC)
counterparts, performed based on the proposed simulation algorithm in \cite%
{cox}. Furthermore, the proposed numerical RTE solver and MC\ simulation
complexities are compared in terms of computation time.

\item Matlab codes for the developed RTE Solver are presented at the end of
the paper \cite{github}.
\end{itemize}

In this context, the remainder of this paper is organized as follows:
Section II presents the UOWC system model as well as the improved TD-RTE
solver. A numerical application of the derived results is shown in Section
III. Section IV concludes the paper with some future directions.

\section{System Model}

\begin{figure}[h]
\begin{center}
\hspace*{-.5cm}\includegraphics[scale=0.5]{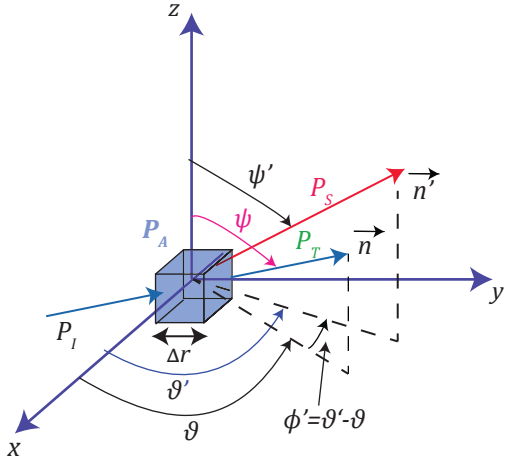} \vspace*{-7.5cm}
\end{center}
\caption{Light propagation in the three-dimensional space.}
\label{sysmod1}
\end{figure}

The light propagation on the three-dimensional space undergoes two main
phenomena: (i) Absorption by which the photons energy is converted into
another form, (ii) and scattering is described as the light interaction with
the medium particles and molecules. As depicted in Fig. \ref{sysmod1}, the
incident power $P_{I}$ , propagating toward the direction $\overrightarrow{n}%
,$ undergoes absorption and scattering in a volume element $\Delta V$ with a
width $\Delta r$, with $P_{A}$ and $P_{S}$ denote the absorbed and scattered
amounts of power within $\Delta V$, respectively. A portion of the incident
power, denoted $P_{T},$ will conserve the propagation direction of the
incident wave. Following the energy-conservation law, one obtains

\begin{equation}
P_{I}=P_{A}+P_{S}+P_{T}.
\end{equation}%
\ The wavelength-dependent absorption and scattering coefficients, measured
in $m^{-1},$ are defined as \cite{survey}%
\begin{equation}
a\left( \lambda \right) =\lim_{\Delta r\rightarrow 0}\frac{P_{A}}{\Delta
rP_{I}},
\end{equation}%
and%
\begin{equation}
b\left( \lambda \right) =\lim_{\Delta r\rightarrow 0}\frac{P_{S}}{\Delta
rP_{I}},
\end{equation}%
where $\lambda $ being the operating wavelength. The total attenuation
coefficient $c\left( \lambda \right) $, measured in $m^{-1},$ is defined as
the sum of the absorption and scattering ones as%
\begin{equation}
c\left( \lambda \right) =a\left( \lambda \right) +b\left( \lambda \right) .
\end{equation}%

In the sequel, we will assume a fixed wavelength. Thus, the absorption,
scattering, and attenuation coefficients are fixed.

\bigskip It is known that the instantaneous light radiance is the solution
of the three-dimensional TD-RTE\ given as \cite{klose}, \cite{mobley}
\begin{align}
\left[ \frac{1}{v}\frac{\partial }{\partial t}+\overrightarrow{n}\cdot
\nabla \right] L(t,r,\vartheta ,\Psi )& =-cL(t,r,\vartheta ,\Psi )+S\left(
t\right)  \notag \\
& +\int_{\vartheta ^{\prime }=0}^{2\pi }\int_{\Psi ^{\prime }=0}^{\pi }\beta
\left( \vartheta ,\Psi ,\vartheta ^{\prime },\Psi ^{\prime }\right)  \notag
\\
& \times L\left( t,r,\vartheta ^{\prime },\Psi ^{\prime }\right) \sin \left(
\Psi ^{\prime }\right) d\Psi ^{\prime }d\vartheta ^{\prime },  \label{RTE}
\end{align}%
where

\begin{itemize}
\item $L(t,r,\vartheta ,\Psi )$ denotes the light radiance at position $r$
from the source, and at time $t$ propagating toward direction $%
\overrightarrow{n}$ in $W/m^{2}/sr.$ It is defined as the amount of power at
distance $r$ and time $t$, per unit of surface and per unit of solid angle.

\item $S\left( t\right) $ being the directed light source radiance at time $%
t $ in $W/m^{3}/sr.$ It is defined as the radiated power density per unit of
volume per unit of solid angle.

\item $\beta \left( {\Psi ,}\vartheta ,{\Psi }^{\prime },\vartheta ^{\prime
}\right) $ is the volume scattering function (VSF), representing the
probability density function (PDF) of the scattered power between two
directions $\overrightarrow{n}$ and $\overrightarrow{n}^{\prime },$
represented by the angles $\left( {\Psi ,}\vartheta \right) $ and $\left( {%
\Psi }^{\prime },\vartheta ^{\prime }\right) ,$ respectively.

\item $\nabla $ is the divergence operator.

\item $v$ being the light celerity in the underwater medium.
\end{itemize}

\begin{Remark}
In the sequel, we will consider the 2D VSF $\beta \left( \vartheta
,\vartheta ^{\prime }\right) $. That is, the RTE\ will be solved in two
dimensions rather than 3D. Consequently, the considered optical radiances
are measured in $W/m^{2}/rad$. In this case, the double integral over the
solid angle of a sphere, given in (\ref{RTE}), will be replaced by a simple
integral of $\vartheta ^{\prime }$ argument over $\left[ 0,2\pi \right] ,$
when $\Psi ^{\prime }=\frac{\pi }{2}.$ Consequently, the 2D RTE\ equation is
expressed as%
\begin{align}
\left[ \frac{1}{v}\frac{\partial }{\partial t}+\overrightarrow{n}\cdot
\nabla \right] L(t,r,\vartheta )& =-cL(t,r,\vartheta )+\int_{0}^{2\pi }\beta
\left( \vartheta ,\vartheta ^{\prime }\right)  \notag \\
& \times L\left( t,r,\vartheta ^{\prime }\right) d\vartheta ^{\prime
}+S\left( t\right) ,  \label{2d}
\end{align}
\end{Remark}

The VSF\ $\beta \left( \vartheta ,\vartheta ^{\prime }\right) $ is related
to the PSF $\widetilde{\beta }\left( \phi ^{\prime }\right) $ as \cite%
{survey}
\begin{equation}
\beta \left( \vartheta ,\vartheta ^{\prime }\right) =b\widetilde{\beta }%
\left( \phi ^{\prime }\right) ,
\end{equation}%
where $\phi ^{\prime }$ denotes the scattering angle between the two
directions $\overrightarrow{n}$ and $\overrightarrow{n}^{\prime }$.

The scalar product of the scattering vector $\overrightarrow{n}^{\prime }$
with unit vectors $\overrightarrow{e}_{x},\overrightarrow{e}_{y}$ is given
by, respectively

\begin{eqnarray}
\overrightarrow{n}^{\prime }\cdot \overrightarrow{e}_{x} &=&\cos \vartheta
^{\prime },  \label{cos} \\
\overrightarrow{n}^{\prime }\cdot \overrightarrow{e}_{y} &=&\sin \vartheta
^{\prime },  \label{sin}
\end{eqnarray}%
where $\vartheta $ and $\vartheta ^{\prime }$ are defined as the angles
between the $x$-axis and the propagation and scattering vectors $%
\overrightarrow{n}$ and $\overrightarrow{n}^{\prime }$ in the $XOY$ plane,
respectively.

\subsection{Single Term HG Function}

A\ popular analytical model for representing anisotropic propagation of
light is the two-dimensions (2D)\ single term Henyey-Greenstein (STHG)
scattering function given as \cite{gao}

\begin{equation}
\widetilde{\beta }_{STHG}\left( g,\phi ^{\prime }\right) =\frac{1-g^{2}}{%
2\pi \left( 1+g^{2}-2g\cos \phi ^{\prime }\right) },0\leq g\leq 1,
\label{sthg}
\end{equation}%
where $g$ accounts for the scattering strength, i.e., isotropic scattering
is defined for $g=0$, while as $g\ $tends to $1$, a peaked scattering
scenario is presented. Interestingly, it has been shown that the value $%
g=0.93$ represents an accurate approximation for the angular distribution of
scattered light in the majority of water types \cite{survey}.

\subsection{Two Terms HG Function}

Due to the inaccurate fitting of the STHG phase function with measurements
at small and large scattering angles, a linear combination of
Henyey-Greenstein phase functions is sometimes used to improve the fit at
small and large angles. The TTHG phase function is given as \cite{kattawar}

\begin{equation}
\widetilde{\beta }_{TTHG}\left( \alpha ,g_{1},g_{2},\phi ^{\prime }\right)
=\alpha \widetilde{\beta }_{STHG}\left( g_{1},\phi ^{\prime }\right)
+(1-\alpha )\widetilde{\beta }_{STHG}\left( g_{2},\phi ^{\prime }\right) ,
\label{tthg}
\end{equation}%
where $\alpha $ is a weighting factor between $0$ and $1$, $g_{1}$ and $%
g_{2} $ stands for scattering strength parameters$.$ It is worth mentioning
that enhanced small-angle scattering is obtained by choosing $g_{1}$ near
one, and enhanced backscatter is obtained by making $g_{2}$ negative.

\subsection{Fournier-Forand Function}

The Fournier-Forand (FF) phase function is among other PSFs\ that have been
proposed as an alternative to the STHG\ and TTHG, in hydraulic optics as
well as in underwater optical environments \cite{mobley}. The two parameters
FF phase function, introduced in \cite{fournierorg}, has a more complex
analytical form compared to its STHG\ and TTHG\ counterparts. Nevertheless,
it depends only on two parameters, and has higher accuracy into modeling
quasi all realistic underwater phase functions. The FF\ PSF\ is expressed as
\cite{fournierorg}, \cite{fournier}%
\begin{align}
\widetilde{\beta }_{FF}\left( \mu ,n_{p},\phi ^{\prime }\right) & =\left[
\begin{array}{l}
\left[ \delta \left( 1-\delta ^{\upsilon }\right) -\upsilon \left( 1-\delta
\right) \right] \sin ^{-2}\left( \frac{\phi ^{\prime }}{2}\right) \\
+\upsilon \left( 1-\delta \right) -\left( 1-\delta ^{\upsilon }\right)%
\end{array}%
\right]  \notag \\
& \times \frac{1}{4\pi (1-\delta ^{2})\delta ^{\upsilon }}+\frac{1-\delta
_{\pi }^{\upsilon }}{16\pi \left( \delta _{\pi }-1\right) \delta _{\pi
}^{\upsilon }}  \notag \\
& \times \left( 3\cos ^{2}\left( \phi ^{\prime }\right) -1\right) ,
\label{ff}
\end{align}
where $n_{p}$ is the real refraction index, and $\mu $ denotes is the slope
parameter of the hyperbolic distribution, with $3\leq \mu \leq 5$, and%
\begin{eqnarray}
\upsilon &=&\frac{3-\mu }{2}, \\
\delta &=&\frac{4}{3(n_{p}-1)^{2}}\sin ^{2}\left( \frac{\phi ^{\prime }}{2}%
\right) ,
\end{eqnarray}%
with $\delta _{\pi \text{ }}$ being $\delta $ evaluated at $\phi ^{\prime
}=\pi .$

\subsection{Optimal Scattering Angles : Improved Algorithm}

\begin{figure}[h]
\begin{center}
\hspace*{-1.1cm}\includegraphics[scale=0.4]{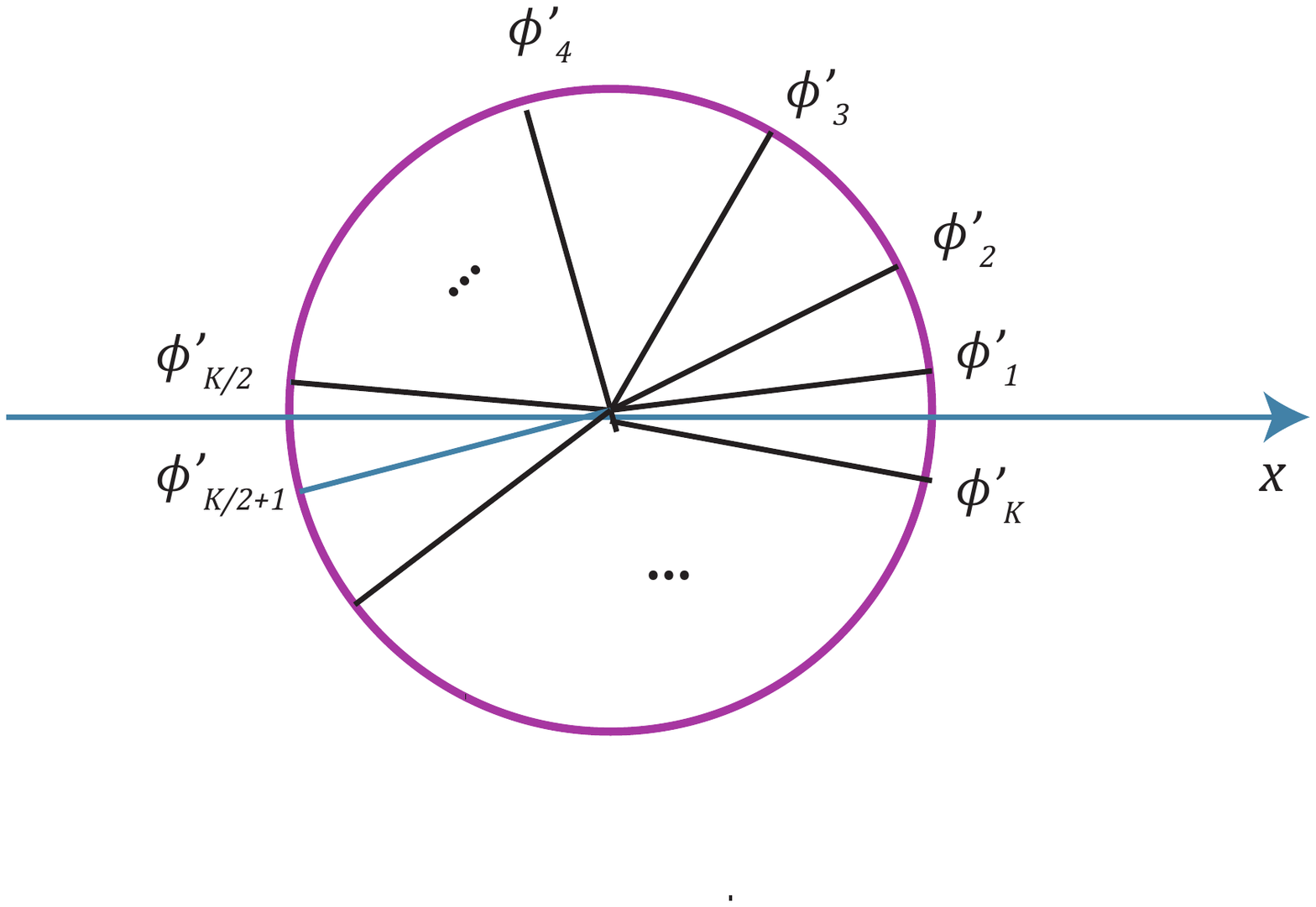} \vspace*{-6.2cm}
\end{center}
\caption{Non-uniform scattering directions discretization.}
\label{sysmod}
\end{figure}

In \cite{gao}, the authors used the uniform discrete ordinate method, based
on discretizing the angular space of propagation into discrete equidistant
directions. However, this approach seems accurate only for isotropic
scattering environments $(g=0)$ and presents some inaccuracies for highly
peaked forward scattering waters \cite{laura}, \cite{alouini}. Considering
the TTHG and FF\ functions in addition to the STHG one, where the angular
directions of light propagation range in the interval $\left[ 0,2\pi \right]
$, the angular space is discretized into $K$ unequally spaced directions $%
\phi _{k}^{\prime },$ as shown in Fig. \ref{sysmod}, minimizing the
following mean squared error \cite{alouini}

\begin{equation}
f(K)=\sum_{i=1}^{K}\int_{d_{i-1}}^{d_{i}}\left( \phi _{i}^{\prime }-\phi
^{\prime }\right) ^{2}\widetilde{\beta }_{X}\left( \Omega ,\phi ^{\prime
}\right) d\phi ^{\prime },
\end{equation}%
where $d_{i}$ denotes the decision thresholds, $X$ denotes either $STHG,$ $%
TTHG,$ or $FF$, while $\left( \Omega ,\phi ^{\prime }\right) $ denotes
either $\left( g,\phi ^{\prime }\right) ,$ $\left( \alpha ,g_{1},g_{2},\phi
^{\prime }\right) ,$ or $\left( \mu ,n_{p},\phi ^{\prime }\right) ,$
respectively.

In order to minimize the MSE of the above function, we must satisfy the two
following conditions

\begin{eqnarray}
d_{i} &=&\frac{\phi _{i}^{\prime }+\phi _{i+1}^{\prime }}{2};1\leq i\leq K,
\label{dii} \\
\phi _{i}^{\prime } &=&\frac{\int_{d_{i-1}}^{d_{i}}\phi ^{\prime }\widetilde{%
\beta }_{X}\left( \Omega ,\phi ^{\prime }\right) d\phi ^{\prime }}{%
\int_{d_{i-1}}^{d_{i}}\widetilde{\beta }_{X}\left( \Omega ,\phi ^{\prime
}\right) d\phi ^{\prime }};1\leq i\leq K.  \label{phii}
\end{eqnarray}

The non-uniform scattering angle discretization process is depicted in
Algorithm 1.

\begin{algorithm}[ht]
\KwData{$ K$, $g$, $\alpha$, $g_1$, $g_2$, $\mu$, $n_p$, $\epsilon$, PSF }
\KwResult{$\phi_k^{\prime}$}
 \Begin{
 \begin{itemize}
\item Initialize $\phi^{\prime} _{k}$ uniformly, i.e., $\phi _{k}^{\prime (0)}=(k-1)\frac{2\pi }{K},1\leq k\leq K$;
 \item Setting $d_{0}=0$ ; \\
 \item $l\gets 0$;\\
 \If{PSF="STHG"}{
    $\Omega \gets g$;
  }
  \ElseIf{PSF="TTHG"}{
   $\Omega \gets \alpha,g_1,g_2$;
  }
  \Else{
   $\Omega \gets \mu,n_p$;
  }
\end{itemize}

  \Repeat{$\phi_k^{\prime(l+1)}-\phi_k^{\prime (l)} < \epsilon$}
     {
    \begin{itemize}
      \item Computing $d_k$ for $1 \leq k \leq K $ using (\ref{dii});\\
 \item Calculating the new values $\phi_k^{\prime (l+1)}$ using (\ref{sthg}),  \\(\ref{tthg}),  (\ref{ff}), and (\ref{phii}), for $1 \leq k \leq K $ ;
\item $l\gets l+1$;
 \end{itemize}
     }
}
\caption{Optimal scattering angles.}
\end{algorithm}

In this paper, the Lloyd-Max algorithm, used in \cite{alouini} and \cite%
{commnet} that provides optimal angles through MSE\ criteria was modified.
The angular discretization in the former version was symmetric with respect
to the reference forward direction $\left( \phi _{1}^{\prime }=0\right) ,$
while in this updated one, the angular discretization is asymmetric with
respect to the $x$-axis. The reference forward direction $\phi _{1}^{\prime
} $ is chosen as $\phi _{1}^{\prime }\neq 0$, which is more practical as the
source beam diverges through a divergent lens, by an initial divergence
half-angle of $\omega .$ That is, $K$ angles will be computed in this
version instead of $\frac{K}{2},$ performed in the former one.

The time coordinate is discretized uniformly into equidistant time instants $%
t_{n}\left( 1\leq n\leq N\right) $, $\Delta t$ denotes the discretization
step between two consecutive time instants $t_{n}$ and $t_{n+1}$, while $N$
accounts for the maximal number of time instants, at which the convergence
of the TD-RTE\ solution is attained.

By discretizing (\ref{cos}) and (\ref{sin}), and plugging them as well as
the time discretization into (\ref{2d}), one obtains

{\normalsize
\begin{align}
\eta _{k}\frac{\partial L_{k}(t_{n},r)}{\partial y}+\xi _{k}\frac{\partial
L_{k}(t_{n},r)}{\partial x}+\frac{1}{v}\frac{\partial L_{k}(t_{n},r)}{%
\partial t}& =-cL_{k}(t_{n},r)  \notag \\
+b\sum_{k_{s}=1}^{K}w_{k,k_{s}}L_{k_{s}}(t_{n},r)+S(t_{n}),k=1,..,K,&
n=1,..,N,  \label{RTEp}
\end{align}%
with}

\begin{itemize}
\item $L_{k}(t_{n},r)$ being the time-dependent radiance at position $r$
propagating toward discrete direction $\phi _{k}^{\prime }$.

\item $\xi _{k}=\cos \phi _{k}^{\prime },$

\item $\eta _{k}=\sin \phi _{k}^{\prime }.$

\item $w_{k,k_{s}}$ denotes the weight terms that substitute the integral
term, with $k$ and $k_{s}$ correspond to the discrete angles of propagation
and scattering directions, respectively.
\end{itemize}

It is worth mentioning that the coefficients $w_{k,k_{s}}$ in the\ equation
above are obtained through quadrature method, detailed in the next
subsection.

\subsection{Accurate Computation of the Integral Term}

In this subsection, in order to solve numerically the integral on the
right-hand side of (\ref{2d}), we incorporate the Simpson's method alongside
with the 5-points and 7-points Boole's rule given by the Newton-Cotes
formulas \cite[Eqs. (25.4.14), (25.4.18)]{abramowitz}, in order to calculate
the weight terms $w_{k,k_{s}}$ given as%
\begin{equation}
w_{1,k_{s}}=\sum\limits_{l=1}^{M}S_{k_{s}}(l);1\leq k_{s}\leq K,
\end{equation}%
where $S_{k_{s}}(l)$ is defined in (\ref{quadra}) at the top of the next
page, $h_{k_{s}}=\frac{\Delta \phi _{k_{s}}^{\prime }}{M}$ denotes an
infinitesimally small quadrature discretization step, $\Delta \phi
_{k_{s}}^{\prime }$ is the difference between the angles $\phi
_{k_{s}}^{\prime }$ and $\phi _{k_{s}+1}^{\prime },$ and $M$ is the number
of discrete points within this area, assumed to be the same for all
sub-intervals of scattering angles. $\widetilde{\beta }_{X}\left( \Omega
,\phi ^{\prime }\right) $ is the PDF of the scattered photons, its
integration over $2\pi $ equals $1.$ Then, it follows from the equation
above that all the terms $w_{1,k_{s}}$ should be normalized by $%
\sum\limits_{k_{s}=1}^{K}w_{1,k_{s}}.$ The remaining terms $w_{k,k_{s}}$ can
be calculated using the formula \cite{alouini}
\begin{figure*}[t]
{\normalsize 
\setcounter{mytempeqncnt}{\value{equation}}
\setcounter{equation}{19} }
\par
{\small
\begin{equation}
\hspace*{-.8cm} S_{k_{s}}(l)=%
\begin{cases}
\frac{h_{k_{s}}}{18}\left( \widetilde{\beta }_{X}\left( \Omega ,\phi
_{k_{s}}^{\prime }\right) +2\widetilde{\beta }_{X}\left( \Omega ,\phi
_{k_{s}}^{\prime }+h_{k_{s}}\right) \right) ,l=1, \\
\frac{2h_{k_{s}}}{36}\left( \widetilde{\beta }_{X}\left( \Omega ,\phi
_{k_{s}}^{\prime }\right) +4\widetilde{\beta }_{X}\left( \Omega ,\phi
_{k_{s}}^{\prime }+h_{k_{s}}\right) +\widetilde{\beta }_{X}\left( \Omega
,\phi _{k_{s}}^{\prime }+2h_{k_{s}}\right) \right) ,l=2 \\
\frac{4h_{k_{s}}}{180}\left(
\begin{array}{l}
7\widetilde{\beta }_{X}\left( \Omega ,\phi _{k_{s}}^{\prime }\right) +32%
\widetilde{\beta }_{X}\left( \Omega ,\phi _{k_{s}}^{\prime
}+h_{k_{s}}\right) +12\widetilde{\beta }_{X}\left( \Omega ,\phi
_{k_{s}}^{\prime }+2h_{k_{s}}\right) \\
+32\widetilde{\beta }_{X}\left( \Omega ,\phi _{k_{s}}^{\prime
}+3h_{k_{s}}\right) +7\widetilde{\beta }_{X}\left( \Omega ,\phi
_{k_{s}}^{\prime }+4h_{k_{s}}\right)%
\end{array}%
\right) ,l=3 \\
\frac{6h_{k_{s}}}{5040}\left(
\begin{array}{l}
41\widetilde{\beta }_{X}\left( \Omega ,\phi _{k_{s}}^{\prime
}+(l-4)h_{k_{s}}\right) +216\widetilde{\beta }_{X}\left( \Omega ,\phi
_{k_{s}}^{\prime }+(l-3)h_{k_{s}}\right) +27\widetilde{\beta }_{X}\left(
\Omega ,\phi _{k_{s}}^{\prime }+(l-2)h_{k_{s}}\right) \\
+272\widetilde{\beta }_{X}\left( \Omega ,\phi _{k_{s}}^{\prime
}+(l-1)h_{k_{s}}\right) +27\widetilde{\beta }_{X}\left( \Omega ,\phi
_{k_{s}}^{\prime }+lh_{k_{s}}\right) +216\widetilde{\beta }_{X}\left( \Omega
,\phi _{k_{s}}^{\prime }+(l+1)h_{k_{s}}\right) \\
+41\widetilde{\beta }_{X}\left( \Omega ,\phi _{k_{s}}^{\prime
}+(l+2)h_{k_{s}}\right)%
\end{array}%
\right) ;4\leq l\leq M-3 \\
\frac{4h_{k_{s}}}{180}\left(
\begin{array}{l}
7\widetilde{\beta }\left( \Omega ,\phi _{k_{s}+1}^{\prime
}-4h_{k_{s}}\right) +32\widetilde{\beta }\left( \Omega ,\phi
_{k_{s}+1}^{\prime }-3h_{k_{s}}\right) +12\widetilde{\beta }\left( \Omega
,\phi _{k_{s}+1}^{\prime }-2h_{k_{s}}\right) \\
+32\widetilde{\beta }\left( \Omega ,\phi _{k_{s}+1}^{\prime
}-h_{k_{s}}\right) +7\widetilde{\beta }\left( \Omega ,\phi
_{k_{s}+1}^{\prime }\right)%
\end{array}%
\right) ,l=M-2 \\
\frac{2h_{k_{s}}}{36}\left( \widetilde{\beta }_{X}\left( \Omega ,\phi
_{k_{s}+1}^{\prime }-2h_{k_{s}}\right) +4\widetilde{\beta }_{X}\left( \Omega
,\phi _{k_{s}+1}^{\prime }-h_{k_{s}}\right) +\widetilde{\beta }_{X}\left(
\Omega ,\phi _{k_{s}+1}^{\prime }\right) \right) ,l=M-1 \\
\frac{h_{k_{s}}}{18}\left( \widetilde{\beta }_{X}\left( \Omega ,\phi
_{k_{s}+1}^{\prime }-h_{k_{s}}\right) +2\widetilde{\beta }_{X}\left( \Omega
,\phi _{k_{s}+1}^{\prime }\right) \right) ,l=M.%
\end{cases}
\label{quadra}
\end{equation}
}
\par
{\normalsize 
\hrulefill 
\vspace*{4pt} }
\end{figure*}
\begin{equation}
w_{k,k_{s}}=w_{1,|k-k_{s}|+1},  \label{symet}
\end{equation}%
\begin{figure}[tbp]
\begin{center}
\hspace*{-1cm}\vspace*{-1cm} \includegraphics[scale=0.46]{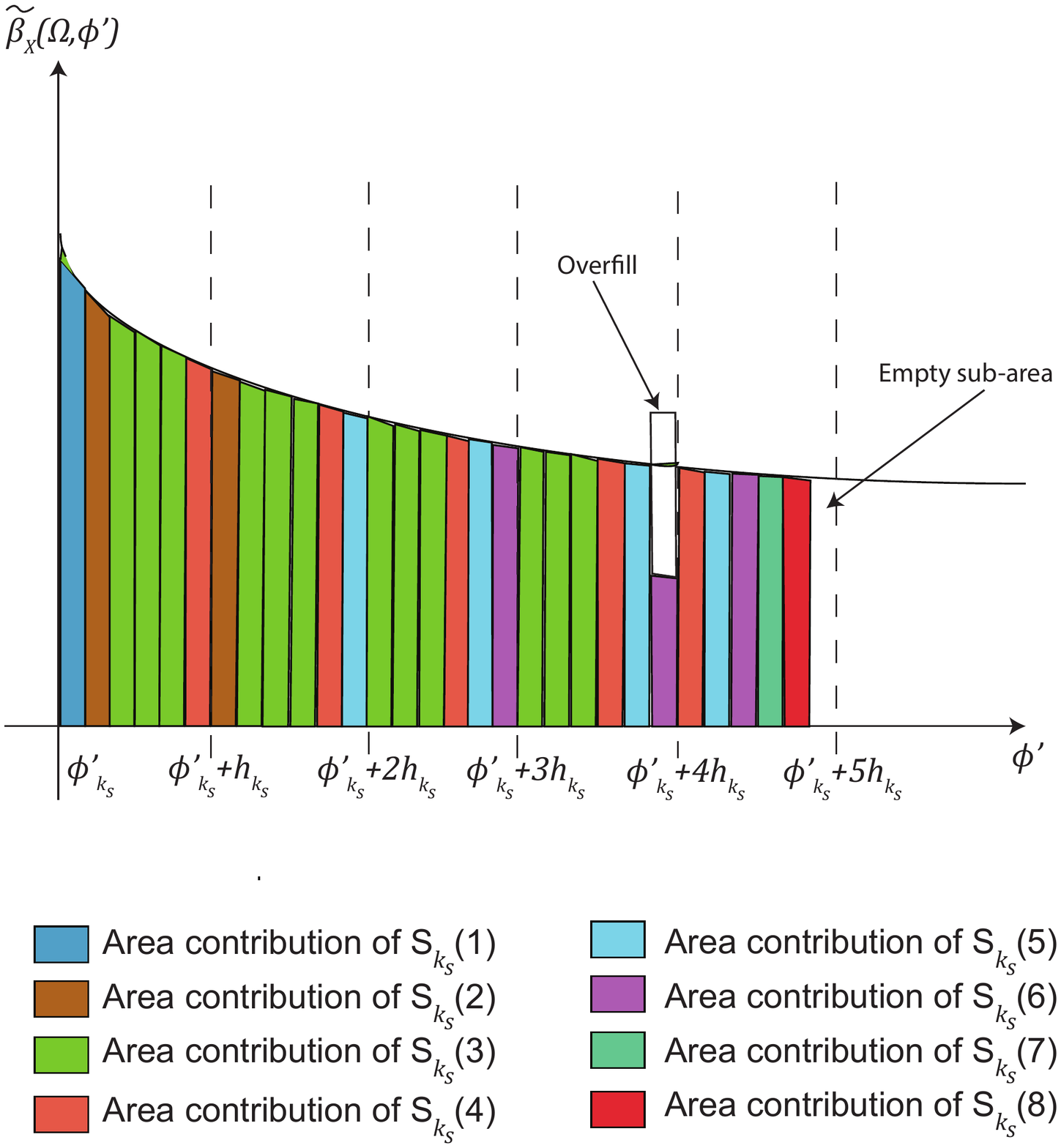}
\end{center}
\caption{Areas rates for the adopted quadrature scheme, using the
computation on 2, 3, 5, and 7 points calculated at terms $%
S_{k_{s}}(l);l=1,...,M$.}
\label{areas}
\end{figure}
\begin{figure}[tbp]
\begin{center}
\hspace*{-0.3cm}\vspace*{-1cm} \includegraphics[scale=0.46]{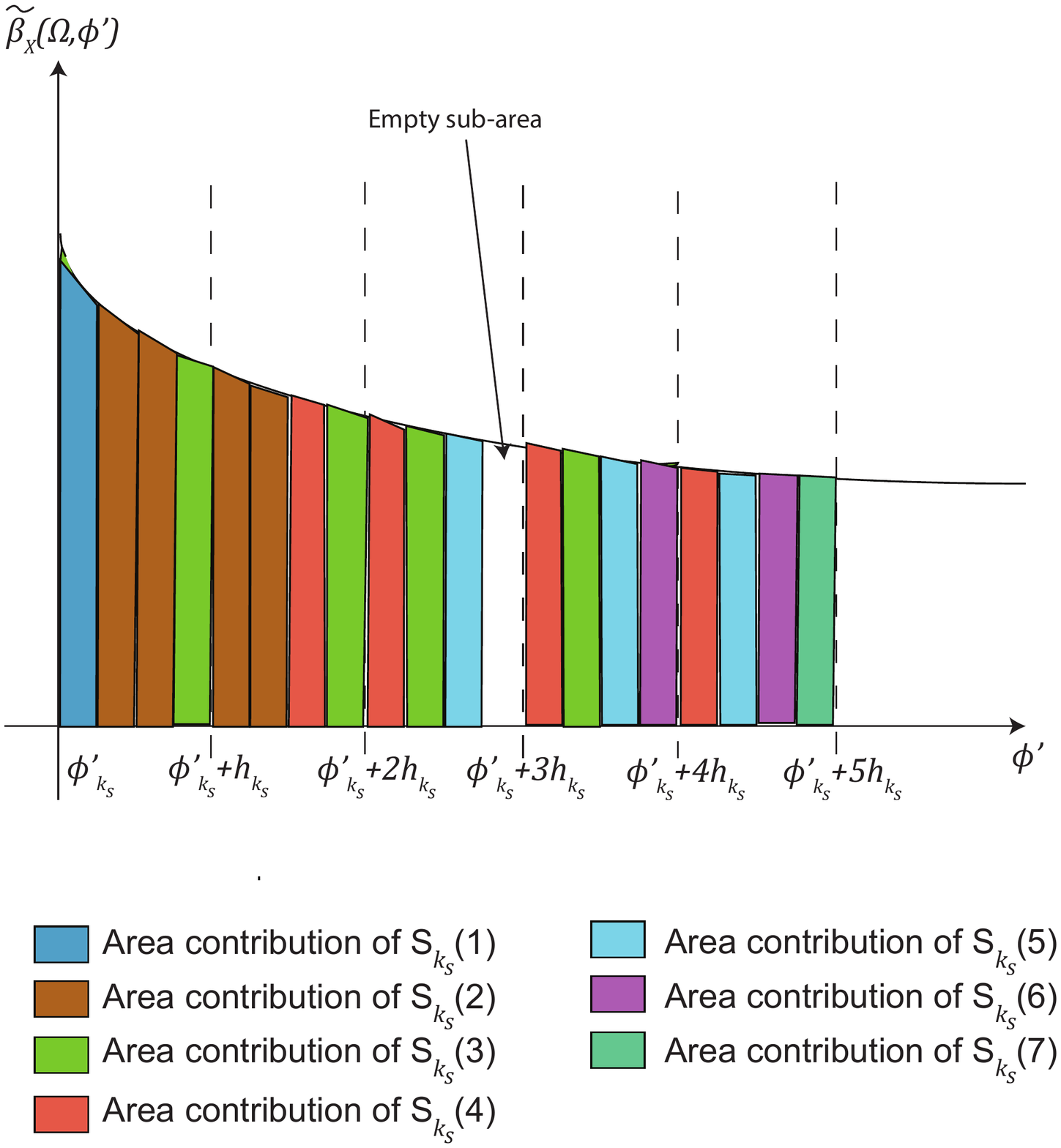}
\end{center}
\caption{Areas rates for the 5-points quadrature scheme used in \protect\cite%
{commnet}, using the computation on 2, 3, and 5 points calculated at terms $%
S_{k_{s}}(l);l=1,...,M$.}
\label{fill22}
\end{figure}

One can remark clearly from (\ref{quadra}) that the quadrature terms
involving 5 points (i.e., $S_{k_{s}}(3)$ and $S_{k_{s}}(M-2))$ were scaled
by $1/2$ from the original equation \cite[Eq. (25.4.14)]{abramowitz}, while
the remaining terms involving two, five and seven points were scaled by a
factor of $1/6$ from \cite[Eqs. (25.4.1)-(25.4.16)]{abramowitz}. In fact,
the area delimited by the function's curve and the $\phi ^{\prime }-$axis
was divided in successive areas. Indeed, as represented in Fig. \ref{areas},
the area $\mathcal{A}_{l}$ delimited by the interval $\left[ \phi
_{k_{s}}^{\prime }+lh_{k_{s}},\phi _{k_{s}}^{\prime }+(l+1)h_{k_{s}}\right]
, $ $5\leq $ $l\leq M-6$ will be recomputed \ in the case of 7-points'
interpolation by three terms on the left, i.e., $S_{k_{s}}(l-1)$, $%
S_{k_{s}}(l)$ and $S_{k_{s}}(l+1)$ as well as by three terms on the right
namely, $S_{k_{s}}(l+2),$ $S_{k_{s}}(l+3)$, and $S_{k_{s}}(l+4).$ Therefore,
each of these terms is scaled by the same factor $1/6$ to get the exact area
rate. Explicitly, we have%
\begin{equation}
\mathcal{A}_{l}\mathcal{=}\sum\limits_{i=0}^{5}S_{k_{s}}\left( l-1+i\right)
,5\leq l\leq M-6.
\end{equation}%
Besides, it can be noticed that the areas $\left[ \phi _{k_{s}}^{\prime
},\phi _{k_{s}}^{\prime }+h_{k_{s}}\right] $ and $\left[ \phi
_{k_{s}}^{\prime }+h_{k_{s}},\phi _{k_{s}}^{\prime }+2h_{k_{s}}\right] $ are
filled through the terms $\left(
S_{k_{s}}(1),S_{k_{s}}(2),S_{k_{s}}(3),S_{k_{s}}(4)\right) $ and $\left(
S_{k_{s}}(2),S_{k_{s}}(3),S_{k_{s}}(4)\right. $, $\left. S_{k_{s}}(5)\right)
$, with scale factors $\left( \frac{1}{6},\frac{1}{6},\frac{1}{2},\frac{1}{6}%
\right) $ and $\left( \frac{1}{6},\frac{1}{2},\frac{1}{6},\frac{1}{6}\right)
$, respectively. Nevertheless, there is an overfill by $\frac{1}{6}$ of the
area associated to the interval $\left[ \phi _{k_{s}}^{\prime
}+lh_{k_{s}},\phi _{k_{s}}^{\prime }+(l+1)h_{k_{s}}\right] ,$for $l=3,M-5,$
as indicated in the rectangle labeled "overfill" within the interval $\left[
\phi _{k_{s}}^{\prime }+3h_{k_{s}},\phi _{k_{s}}^{\prime }+4h_{k_{s}}\right]
$ in Fig. \ref{areas}, as well as $\frac{1}{6}$ of the area $\left[ \phi
_{k_{s}}^{\prime }+lh_{k_{s}},\phi _{k_{s}}^{\prime }+(l+1)h_{k_{s}}\right]
, $ for $l=4,M-6,$ not filled as represented by the blank bar within the
interval $\left[ \phi _{k_{s}}^{\prime }+4h_{k_{s}},\phi _{k_{s}}^{\prime
}+5h_{k_{s}}\right] $ in the same figure, within the whole area $\left[ \phi
_{k_{s}}^{\prime },\phi _{k_{s}+1}^{\prime }\right] .$ Interestingly, the
5-points scheme developed in \cite{commnet} and presented in Fig. \ref%
{fill22} is based on computing the areas $\left[ \phi _{k_{s}}^{\prime
}+lh_{k_{s}},\phi _{k_{s}}^{\prime }+(l+1)h_{k_{s}}\right] $ $(l=1,..,M-2)$
using the 2, 3, and 5-points computation, with corresponding scales of $%
\frac{1}{4},$ $\frac{1}{2},$ and $\frac{1}{4}$, respectively, so as to fill
the whole area delimited by the $\phi ^{^{\prime }}$ axis and the function's
curve.

\begin{Remark}
It is worthy to mention that this quadrature scheme was applied up only to
5-points quadrature scheme equivalently in \cite{commnet} as depicted in
Fig. \ref{fill22}, by considering $\left[ \phi _{1}^{\prime },\phi
_{K}^{\prime }\right] $ as the surface of interest, and the points $\phi
_{k_{s}}^{\prime }(1\leq k_{s}\leq K)$ as the integration points. However,
this quadrature method is inaccurate considering this scheme, since the
non-uniform discretization step $\Delta \phi _{k_{s}}^{\prime }$ of each
interval is relatively great (e.g., $\Delta \phi _{1}^{\prime }=6%
{{}^\circ}%
)$, and consequently, that applied method is not valid.
\end{Remark}

\subsection{Finite Difference Equation}

As a third step of the process, and in order to solve the spatial derivative
terms in (\ref{RTEp}), the upwind finite difference equation is involved.

The area between the transmitter and the receiver is divided into $I$ grid
points horizontally and $J$ grid points vertically. It is noteworthy that in
order to improve the computation accuracy of the upwind finite difference
scheme used in \cite{klose}, we involve one more point in each formula.

The Taylor-series development of the radiance function $L_{i,j,k}(t_{n})$
near the four neighbor points $(i\pm 1,j),(i\pm 2,j)$ close to $(i,j),$ with
$i=1,...,I$ and $j=1,...,J,$ is given by

\begin{eqnarray}
L_{i\pm 1,j,k}(t_{n}) &\approx &L_{i,j,k}(t_{n})\pm \Delta y\frac{\partial
L_{i,j,k}(t_{n})}{\partial y}, \\
L_{i\pm 2,j,k}(t_{n}) &\approx &L_{i,j,k}(t_{n})\pm 2\Delta y\frac{\partial
L_{i,j,k}(t_{n})}{\partial y}.
\end{eqnarray}

In a similar manner, the same development is performed at the points $%
(i,j\pm 1),(i,j\pm 2).\qquad \qquad $

By performing some algebraic manipulation on the equations above, and using
the same reasoning likewise for the other cases of $\xi _{k},\eta _{k}$,
improved finite difference formulas are obtained as{\normalsize
\begin{equation}
\frac{\partial L_{i,j,k}(t_{n})}{\partial y} \approx \left\{
\begin{array}{c}
\frac{2L_{i,j,k}(t_{n})-L_{i-1,j,k}(t_{n})-L_{i-2,j,k}(t_{n})}{3\Delta y}%
,\eta _{k}>0 \\
\frac{L_{i+1,j,k}(t_{n})+L_{i+2,j,k}(t_{n})-2L_{i,j,k}(t_{n})}{3\Delta y}%
,\eta _{k}<0%
\end{array}%
\right.  \label{deriv1}
\end{equation}
\begin{equation}
\frac{\partial L_{i,j,k}(t_{n})}{\partial x} \approx \left\{
\begin{array}{c}
\frac{2L_{i,j,k}(t_{n})-L_{i,j-1,k}(t_{n})-L_{i,j-2,k}(t_{n})}{3\Delta x}%
,\xi _{k}>0 \\
\frac{L_{i,j+1,k}(t_{n})+L_{i,j+2,k}(t_{n})-2L_{i,j,k}(t_{n})}{3\Delta x}%
,\xi _{k}<0%
\end{array}%
\right. ,  \label{deriv2}
\end{equation}
}where $\Delta x$ and $\Delta y$ stand for the discretization steps in the $%
x $ and $y$ axes, respectively.

One can ascertain that each partial derivative is associated to two formulas
depending on the sign of $\xi _{k}$ and $\eta _{k}.$

Regarding the time derivative, the forward Euler difference formula was used
as\cite{klose}
\begin{equation}
\frac{\partial L_{i,j,k}(t_{n})}{\partial t}=\frac{%
L_{i,j,k}(t_{n+1})-L_{i,j,k}(t_{n})}{\Delta t};n=1,..,N-1,  \label{deriv3}
\end{equation}%
with $\Delta t$ being the discretization step for the time coordinate.

By plugging the partial derivatives (\ref{deriv1}), (\ref{deriv2}), and (\ref%
{deriv3}) into (\ref{RTEp}), and performing some manipulations, we obtain
the recursive equation (\ref{solution}) given at the top of the next page.
\begin{figure*}[t]
{\normalsize 
\setcounter{mytempeqncnt}{\value{equation}}
\setcounter{equation}{27} }
\par
\begin{align}
L_{i,j,k}(t_{n+1})& =L_{i,j,k}(t_{n})\left[ 1-cv\Delta t\pm \frac{2\eta
_{k}v\Delta t}{3\Delta y}\pm \frac{2\xi _{k}v\Delta t}{3\Delta x}\right] \mp
\eta _{k}v\Delta t\frac{L_{i\pm 1,j,k}(t_{n})+L_{i\pm 2,j,k}(t_{n})}{3\Delta
y}  \notag \\
& \mp \xi _{k}v\Delta t\frac{L_{i,j\pm 1,k}(t_{n})+L_{i,j\pm 2,k}(t_{n})}{%
3\Delta x}+bv\Delta
t\sum_{k_{s}=1}^{K}w_{k,k_{s}}L_{i,j,k_{s}}(t_{n})+v\Delta tS_{0}.
\label{solution}
\end{align}%
\par
{\normalsize 
\hrulefill 
\vspace*{4pt} }
\end{figure*}
It is noteworthy that the above equation depicts the recursive numerical
solution of the proposed TD-RTE\ solver for the instantaneous light
radiance, in terms of the system and channel parameters, namely the source
radiance, the discretization steps in space and time coordinates, the number
of directions, scattering and absorption coefficient, and light celerity in
the medium as well.
\begin{figure}[!tbp]
\begin{center}
\hspace*{-1cm}\includegraphics[scale=0.45]{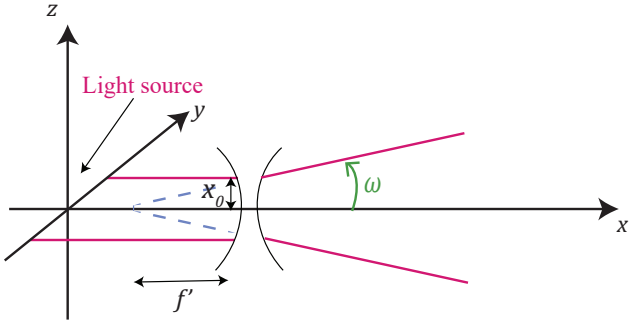}\vspace*{-7cm}
\end{center}
\caption{Collimated beam divergence through a divergent lens.}
\label{lens}
\end{figure}

Without loss of generality, we consider a point source with constant power
over time $S_{0},$ defined at a specific point in the transmitter plane, $%
\left( \text{i.e., }1,\frac{I-1}{2}+1\right) $. In practice, the total
transmit power is radiated in the form of a collimated beam, through a
divergent optical lens with a certain focal length $f^{\prime }$ with an
associated divergence half-angle $\omega ,$ so as to produce a divergent
beam \cite{cox}. $f^{\prime }$ and $\omega $ are related as

\begin{equation}
f^{\prime }=-\frac{x_{0}}{\omega },
\end{equation}
with $x_{0}$ being the beam waist radius at the lens as shown in Fig. \ref%
{lens}. In this case, the source radiance equals to the ratio between the
source power and a circular surface of radius $x_{0}$ and to the divergence
angle $2\omega $.

\begin{Remark}
It is worthy to mention that by neglecting the radiance variations over time
$\left( i.e.,\frac{\partial L_{i,j,k}(t)}{\partial t}=0\right) ,$ the light
radiance $L_{i,j,k}(t_{n})$ becomes time-independent. That is, the TD-RTE\
equation in (\ref{solution}) becomes TI-RTE expressed as%
\begin{align}
L_{i,j,k}^{(l+1)} &=\left( \frac{1}{\mp \frac{2\eta _{k}}{3\Delta y}\mp
\frac{2\xi _{k}}{3\Delta x}+c}\right) \left( \mp \eta _{k}\frac{L_{i\pm
1,j,k}^{(l)}+L_{i\pm 2,j,k}^{(l)}}{3\Delta y}\right.  \notag \\
&\left. \mp \xi _{k}\frac{L_{i,j\pm 1,k}^{(l)}+L_{i,j\pm 2,k}^{(l)}}{3\Delta
x}+b\sum_{k_{s}=1}^{K}w_{k,k_{s}}L_{i,j,k_{s}}^{(l)}+S_{0}\right) ,
\label{TI}
\end{align}%
where $l$ denotes the solution iteration index.
\end{Remark}

\subsection{Received Power Calculation}

In our analysis, the receiver is placed on the $YOZ$ plane. Knowing that the
calculated radiance is performed in the $XOY$ plane, the received power is
calculated by summing up the light radiance at grid points in the receiver
plane perpendicular to the $x-$axis (i.e., $YOZ$ plane). Without loss of
generality, we assume that the receiver aperture placed at the receiver
plane is divided into $\frac{\Delta y}{2}-$equidistant circular surfaces,
being defined as \cite{alouini}

\begin{equation}
A_{l}=\left\{
\begin{array}{l}
\pi \left( \frac{\Delta y}{2}\right) ^{2},l=1 \\
2\pi \left( \Delta y\right) ^{2}\left( l-1\right) ,2\leq l\leq L%
\end{array}%
\right. ,  \label{surf}
\end{equation}%
where $L$ denotes the number of the circular surfaces within the receiver
aperture given in terms of the receiver aperture of radius $R$
\begin{equation}
L=\frac{R}{\Delta y}.
\end{equation}

Let $\Delta \phi _{p}^{\prime }$, with $1\leq p\leq P,$ denotes the
difference between two directions $\phi _{p}^{\prime }$ and $\phi
_{p+1}^{\prime }$ within the receiver FOV in the $XOY$ plane, discretized in
the same way as the scattering angles following equation (\ref{dii}) and (%
\ref{phii}), with $P$ denotes the total number of discrete directions within
the receiver FOV. Without loss of generality, the radiance within the angle
interval delimited by $\Delta \phi _{p}^{\prime }$ in $XOY\ $plane, is
assumed to be constant. By considering also symmetric scattering in the
elevation direction, the surface power density is uniform within each
elementary circular surface within the receiver aperture. Therefore, the
total received power can be expressed as \cite{alouini}

\begin{equation}
P_{r}(t_{n})=\sum_{l=1}^{L}A_{l}\sum_{p=1}^{P}\Delta \phi _{p}^{\prime
}L_{l+\left( \frac{I-1}{2}\right) ,J,p}(t_{n}),n=1,...,N,  \label{power}
\end{equation}

\begin{algorithm}[h]
\KwData{$ K$, $g$, $\alpha$, $g_1$, $g_2$, $\mu$, $n_p$, $\epsilon$, PSF, $\Delta x$, $\Delta y$, $\Delta t$, $t_N$, $x_{max}$, $y_{max}$, $R$, $S_0$, $FOV$, $P$ }
\KwResult{$L_{i,j,k}(t_n)$,$P_r(t_n)$}
 \Begin{
 \begin{itemize}
 \item Discretize $\phi^{\prime}_k$ with non-uniform distribution \\$\phi^{\prime}_k$= \textbf{Optimal scattering angles} $\left( K, g, \alpha, g_1, g_2, \mu, n_p, \epsilon, PSF \right)$,\\ $1 \leq k \leq K $;
 \item Computing the quadrature terms $w_{k,k_s}$ \\ for $k,k_s \in [1,K]$ using (\ref{quadra}) and (\ref{symet});
 \item $I\gets  \floor{\frac{y_{max}}{\Delta y}}+1$;
 \item $J\gets \floor{\frac{x_{max}}{\Delta x}}+1$;
 \item  $\phi^{\prime}_p\gets \phi^{\prime}_k$ for $\phi^{\prime}_k \in FOV, k=1,..,K, p=1,..,P$;
 \item $L\gets \frac{R}{\Delta y}$;
  \item $N\gets \floor{\frac{t_{N}}{\Delta_t}}+1$;
 \item Calculate iteratively the radiance $L_{i,j,k}(t_n)$ \\using (\ref{solution}) for $i=1,..,I$, $j=1,..,J$, $k=1,..,K$,\\ and $n=1,..,N$;
 \item Evaluate the total received power $P_r(t_n)$, \\using (\ref{power}) ;
 \end{itemize}
     }
\caption{RTE power computation.}
\label{alg2}
\end{algorithm}
It is worthy to mention that the Matlab code of Algorithm \ref{alg2} for the
RTE resolution and power computation is presented in Appendix A, as well as
in \cite{github}.

\section{BER performance of UOWC}

In this section, and based on the derived results, we investigate the bit
error rate (BER) performance of the underwater optical wireless
communication system subject to absorption and scattering, in terms of
channel and system parameters.

\subsection{Signal-to-Noise Ratio}

The instantaneous signal-to-noise ratio (SNR) at the output of a receiver
employing the direct detection technique, in the presence of thermal and
shot noise, is expressed as \cite{commnet}, \cite{fso1}, \cite{anguita}

\begin{align}
SNR\left( t_{n}\right) &=\frac{I_{P}^{2}(t_{n})}{N_{0}}=\frac{\left(
R_{s}FP_{r}\left( t_{n}\right) \right) ^{2}}{\sigma _{s}^{2}+\sigma _{th}^{2}%
}, \\
&=\frac{\left( R_{s}FP_{r}\left( t_{n}\right) \right) ^{2}}{%
2q(I_{P}(t_{n})+I_{D})B_{w}+\frac{4\kappa T_{e}B_{w}}{R_{L}}},  \notag
\end{align}%
where

\begin{itemize}
\item $I_{P}(t_{n}):$ Incident light photo current $(A),$

\item $N_{0}:$ Total noise power ($W),$

\item $R_{s}:$ Photodetector responsivity ($A/W)$,

\item $F:$ Photodetector gain factor ($F=1$ for PIN photodetector),

\item $\sigma _{s}^{2}:$ Shot noise power ($W),$

\item $\sigma _{th}^{2}:$ Thermal noise power ($W),$

\item $q:$ Electrical elementary charge ($1.6\times 10^{-19}C$),

\item $I_{D}:$ Shot noise dark current $(A),$

\item $B_{w}:$ Electrical filter bandwidth $(Hz)$,

\item $\kappa :$ Boltzmann constant $(J\cdot K^{-1}),$

\item $T_{e}:$ Receiver temperature ($K$),

\item $R_{L}:$ Electrical receiver load resistance ($\Omega ).$
\end{itemize}

\subsection{Bit Error Rate}

The BER of a communication system employing On-Off Keying (OOK) modulation
scheme, with respect to a received SNR is defined as \cite[Eq. (4.24)]{fso1}%
\qquad

\begin{align}
P_{e}\left( t_{n}\right) &=Q\left( \sqrt{SNR(t_{n})}\right)  \notag \\
&=Q\left( \frac{R_{s}FP_{r}\left( t_{n}\right) }{\sqrt{%
2q(I_{P}(t_{n})+I_{D})B_{w}+\frac{4\kappa T_{e}B_{w}}{R_{L}}}}\right) ,
\label{ber}
\end{align}%
with $Q\left( .\right) $ denotes the Gaussian $Q$-function \cite[Eq. (4.1)]%
{alouini-simon}. Note that the values of $R_{s},$ $F,$ and $I_{D}$ can be
usually retrieved from the datasheet of the photodiode product.

\section{Performance and Complexity evaluations}

This section presents the evaluation performance of the proposed TD-RTE\
solver in terms of computation accuracy as well as complexity. The TD-RTE\
results are given in three dimensions (3D) as a function of the propagation
distance as well as time instants, while the TI-RTE\ results are shown in
two dimensions (2D) versus the distance. The respective MATLAB\ codes of the
proposed RTE\ solver are available in \cite{github}. The RTE\ and MC\
simulations complexities are depicted in terms of time consumption per each
computation step. We depict a single scattering scenario per each PSF among
the three considered scattering functions, for fixed values of $g=0.93$ for
the STHG function, $\alpha =0.9832,g_{1}=0.8838,g_{2}=-0.9835$ for the TTHG\
function, while we set $n=1.33,$ $\mu =$ $3.483$ for the Fournier-Forand
scattering model. Two water types are taken into account, namely Harbor-I ($%
b=0.91,$ $c=1.1)$ and Harbor-II ($b=1.8177,$ and $c=2.2)$\ waters with: $%
Y=20 $ cm as the tank altitude, $t_{N}=20$ ns, $\Delta x=5$ cm, $\Delta y=1$
cm, and $\Delta t=25$ ps as the discretization steps in the $x$-axis, $y$%
-axis, and time coordinate, respectively, and $K=22$ as the number of
discrete directions. Additionally, we considered a transmitting optical lens
with a beam waist radius of $x_{0}=1$ mm. The values of $R_{s} $, $I_{D}$,
and $T_{e}$ were taken from \cite{anguita}. The respective Monte Carlo
results were performed based on the simulations provided in \cite{cox}. In
this regard, the CDF\ numerical computation in this latter was modified in
order to evaluate the CDF with respect to the generated scattering angles
accurately. Furthermore, the number of generated angles was adjusted as a
function of the propagation distance.

Figs. 6-10 depict the TI-RTE normalized received power results over time
versus the distance over Harbor-I and Harbor-II water types, assuming STHG\
scattering function. Actually, determining the number of discrete points $M$
within each interval $[\phi _{k_{s}}^{\prime },\phi _{k_{s}+1}^{\prime }]$
is of paramount importance to improve the accuracy of the proposed RTE\
solver. Particularly, we depict in Figs. 6-8 the RTE solver behavior using
the considered quadrature schemes, namely 3-points, 5-points, and 7-points
schemes. One can remark clearly that the curves corresponding to 3-points
and 5-points schemes shown in Fig. 6, Fig. 7, and Fig. 10 converge to the
exact solution by increasing the number of discretization points $M$ (lower
discretizing step $h_{k_{s}}$) within each surface $[\phi _{k_{s}}^{\prime
},\phi _{k_{s}+1}^{\prime }]$ $($i.e., $M=50$ for 3-points, and $M=40$ for
5-points), while for the 7-points scheme in figures 8 and 10, the
convergence needs a lesser number of points than for the 3-points and
5-points schemes (i.e., $M=7$).

\begin{figure}[h]
\begin{center}
\hspace*{-0.25cm}\includegraphics[scale=0.47]{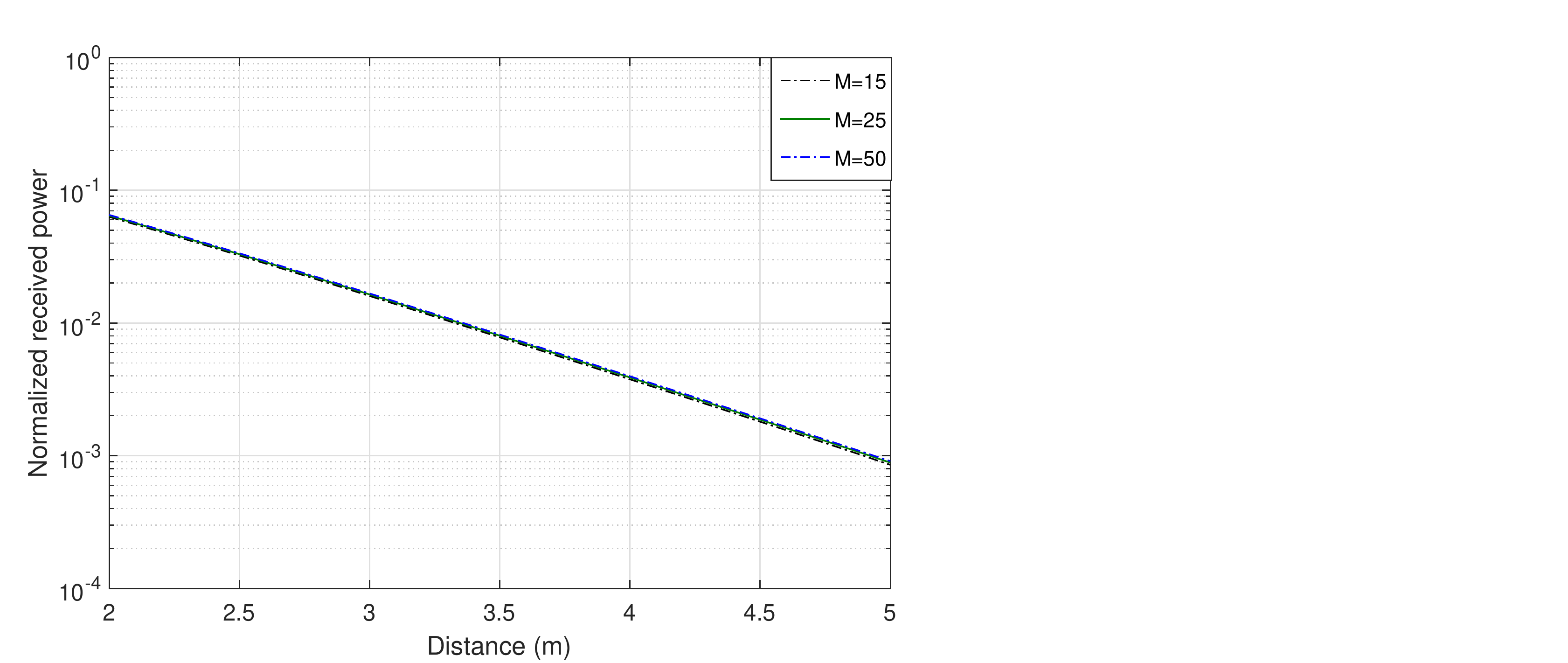}
\end{center}
\caption{Average normalized received power over time propagation distance in
a Harbor-II turbid water medium for the STHG function using 3-pts scheme.}
\label{fig6}
\end{figure}
\begin{figure}[h]
\begin{center}
\hspace*{-0.25cm}\includegraphics[scale=0.47]{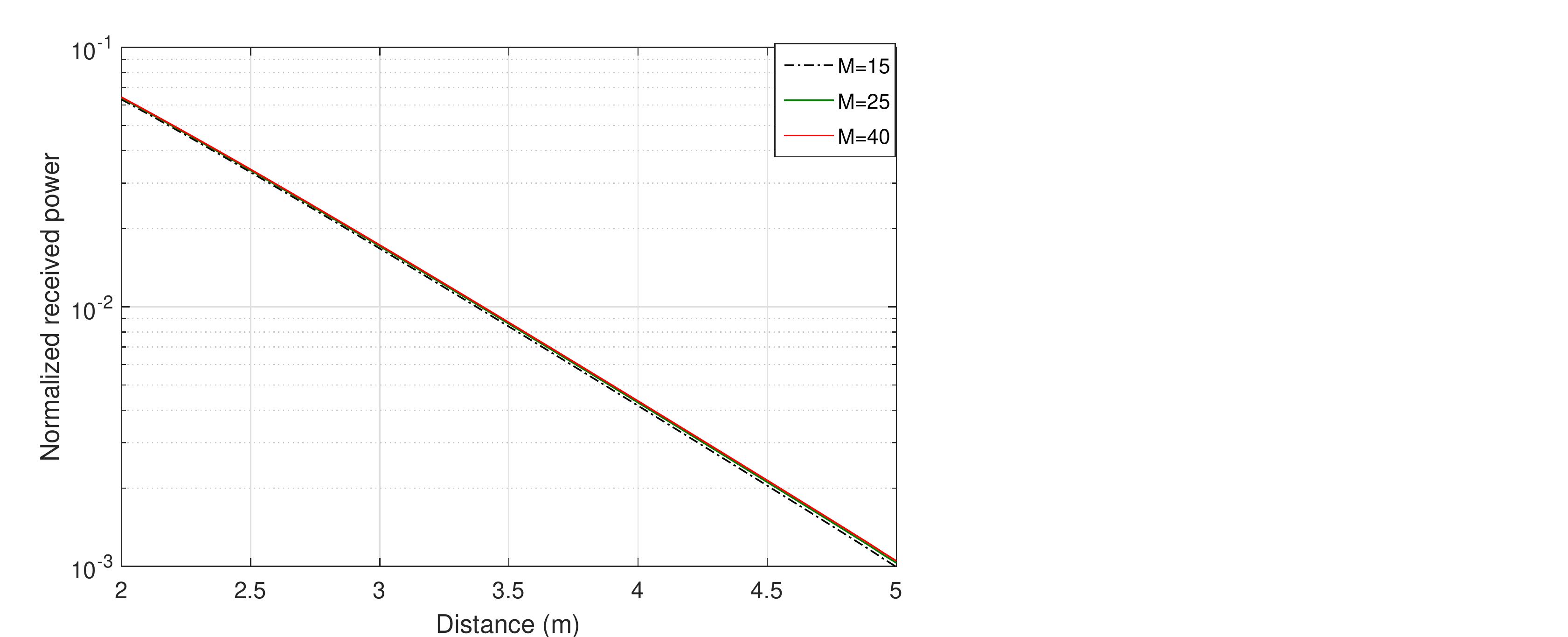}
\end{center}
\caption{Average normalized received power over time propagation distance in
a Harbor-II turbid water medium for the STHG function using 5-pts scheme.}
\label{fig7}
\end{figure}
\begin{figure}[h]
\begin{center}
\hspace*{-1cm}\includegraphics[scale=0.47]{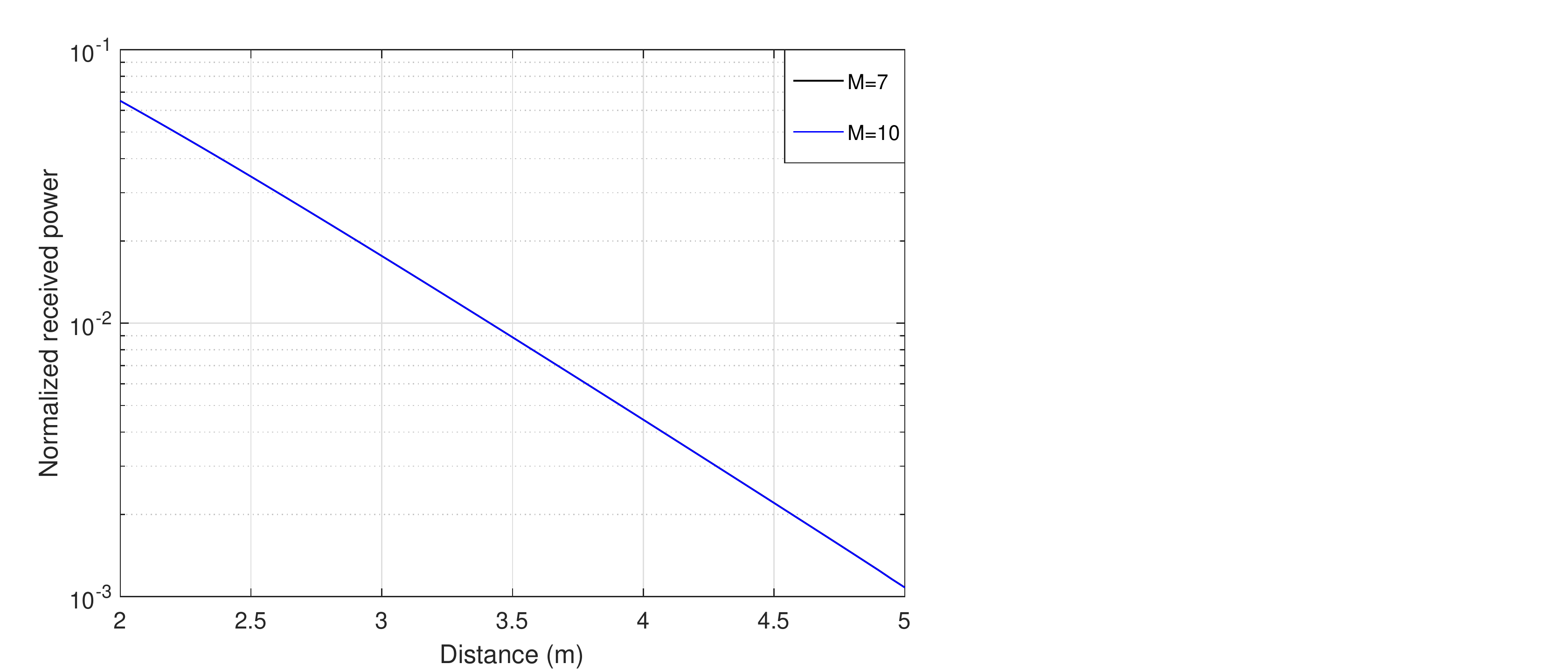}
\end{center}
\caption{Average normalized received power over time propagation distance in
a Harbor-II turbid water medium for the STHG function using 7-pts scheme.}
\label{fig8}
\end{figure}
\begin{figure}[h]
\begin{center}
\hspace*{-1.2cm}\includegraphics[scale=0.47]{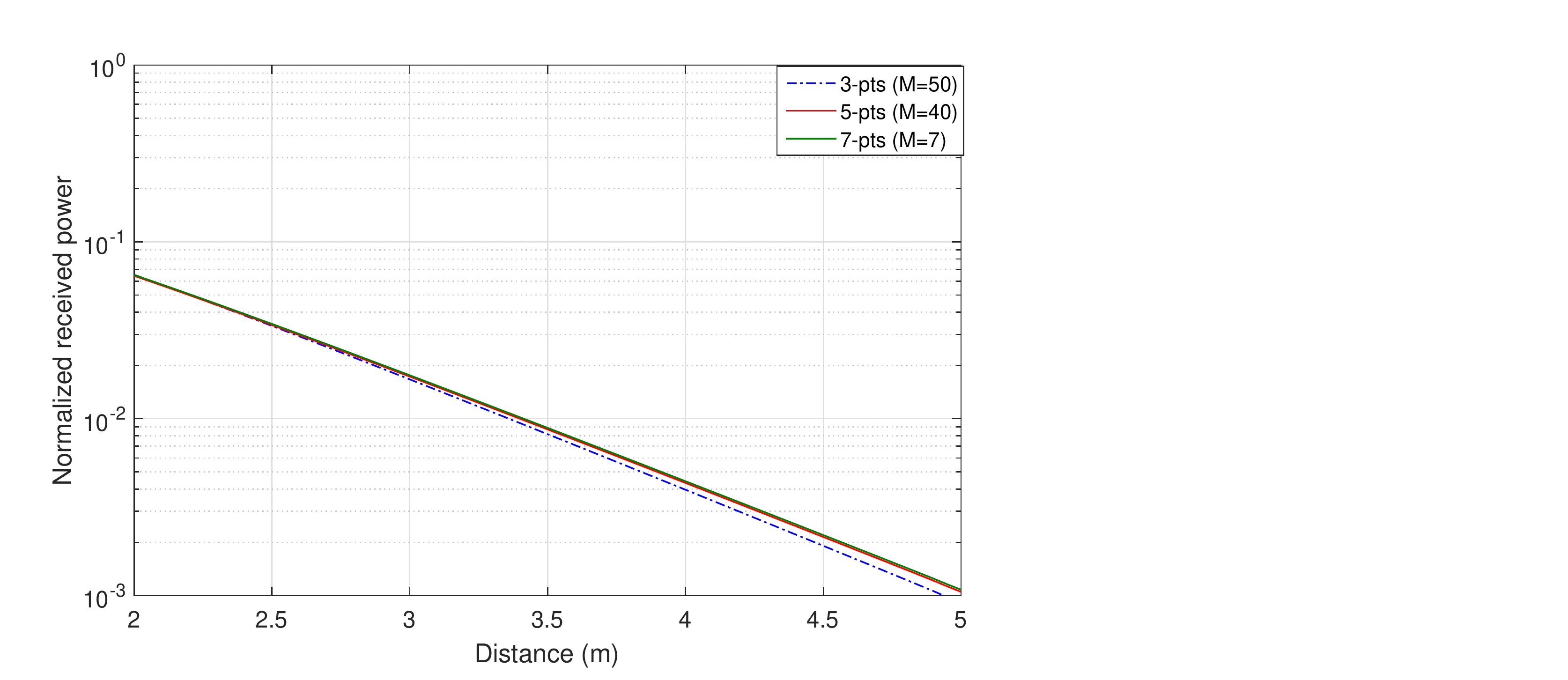}
\end{center}
\caption{Normalized received power versus propagation distance in a
Harbor-II turbid water medium for the STHG function.}
\label{fig9}
\end{figure}
\begin{figure}[h]
\begin{center}
\hspace*{-0.75cm}\includegraphics[scale=0.47]{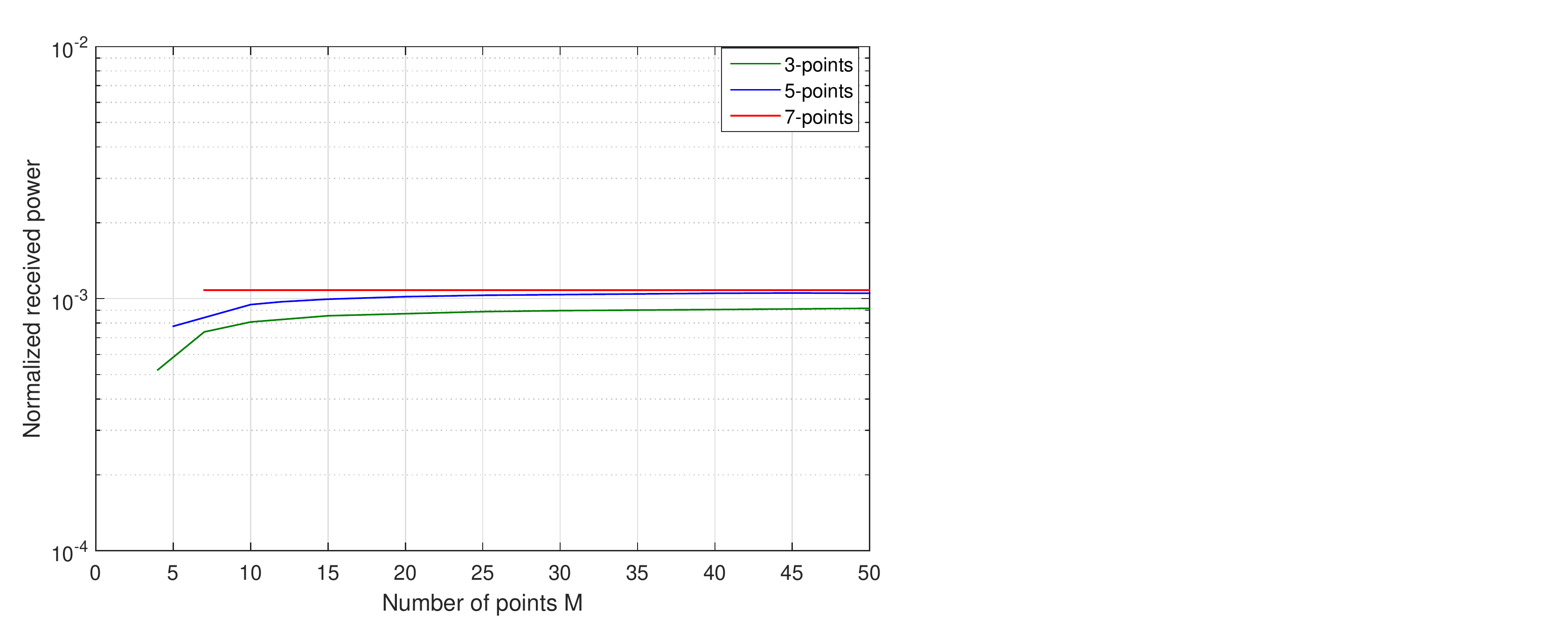}
\end{center}
\caption{Normalized received power versus propagation distance in a
Harbor-II turbid water medium for the STHG function.}
\label{fig10}
\end{figure}

In terms of computation complexity of the adopted quadrature schemes, Table
I shows the evaluation time in seconds of the implemented 7-points
quadrature scheme, for various values of the scattering angles number $K$,
compared with the 3 and 5-points schemes developed in \cite{alouini} and
\cite{commnet}, respectively. These latter were adapted similarly as
performed to the 7-points scheme, where they were applied for
infinitesimally small sub-intervals between two successive scattering
angles. Additionally, the respective number of points $M$ within each
subinterval of two successive scattering angles is taken as $M=7,$ $40,$ and
$50$ for the 3, 5, and 7-points schemes, respectively. Interestingly, it can
be noticed that the 7-points schemes is $8$-$20$ ms less than its 3 and 5
counterparts. In fact, the higher the number of points $M$, the greater the
computation time needed. Furthermore, one can ascertain that the time
consumption increases slightly as a function of $K$ for the aforementioned
schemes. Additionally, it can be obviously seen that the total consumed time
slightly differs between the 3 schemes. In fact, since the quadrature weight
coefficients are calculated once and outside the main loops $(i.e.,$ $%
i=1,...,I;$ $j=1...,J;$ $k=1,...,K;$ $n=1,...,N),$ the computation time is
not impacted significantly.

\begin{table}[t]
\caption{Evaluation time of the numerical integration quadrature schemes (in
seconds).}\centering
\par
\begin{tabular}{c|c|c|c}
\hline
\backslashbox{$K$}{Quadrature Scheme} & 3 points & 5 points & 7 points \\
\hline
$22$ & 0.0139 & 0.0132 & 0.0017 \\ \hline
$26$ & 0.0147 & 0.0134 & 0.0027 \\ \hline
$30$ & 0.0187 & 0.0148 & 0.0052 \\ \hline
$35$ & 0.0175 & 0.0142 & 0.0063 \\ \hline
$40$ & 0.0229 & 0.0151 & 0.0070 \\ \hline
$50$ & 0.0276 & 0.0165 & 0.0078 \\ \hline
\end{tabular}%
\end{table}

Figs. 11-13 depict the TD-RTE normalized received power result versus
distance and time, for Harbor-I, in three dimensions, taking into account
the considered scattering functions (STHG, TTHG, and FF). One can ascertain
that the received power decreases as a function of the distance, i.e., the
farther the communication nodes are, the higher the power path-loss is due
to water attenuation phenomena. Moreover, we ascertain that at initial time
instants, the received power at a given distance is lower initially, and
starts gradually increasing as a function of time until reaching the
convergence level when the received power remains constant in time.
Actually, at initial time instants, few photons reach the receiver plane,
and consequently, it results in lower received power. More photons reach the
receiver side resulting in an increase of the received power versus time.
\begin{figure}[h]
\begin{center}
\hspace{-.5cm} \includegraphics[scale=0.47]{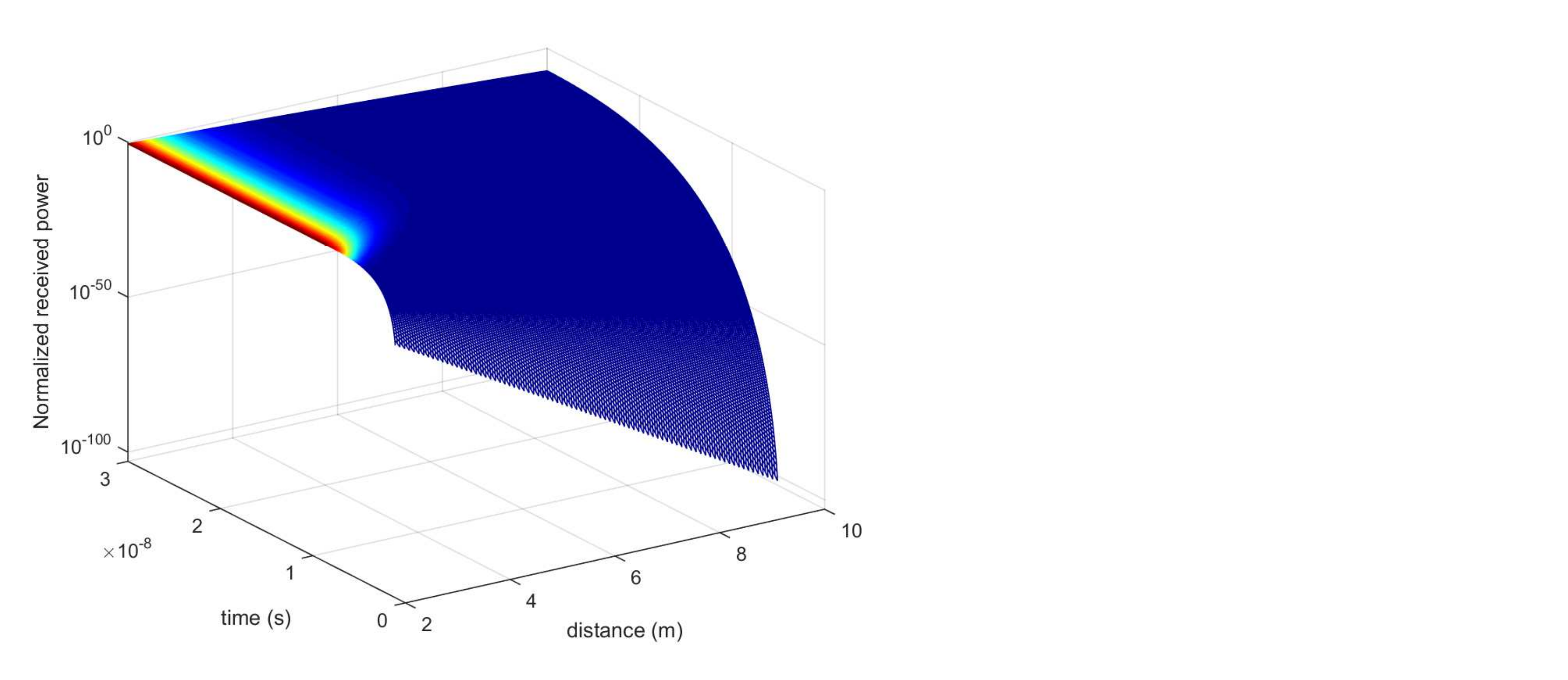}
\end{center}
\caption{Normalized received power in 3D versus distance and time,
considering STHG\ function and a receiver aperture of 10cm.}
\label{fig11}
\end{figure}
\begin{figure}[h]
\begin{center}
\hspace*{-.5cm} \includegraphics[scale=0.47]{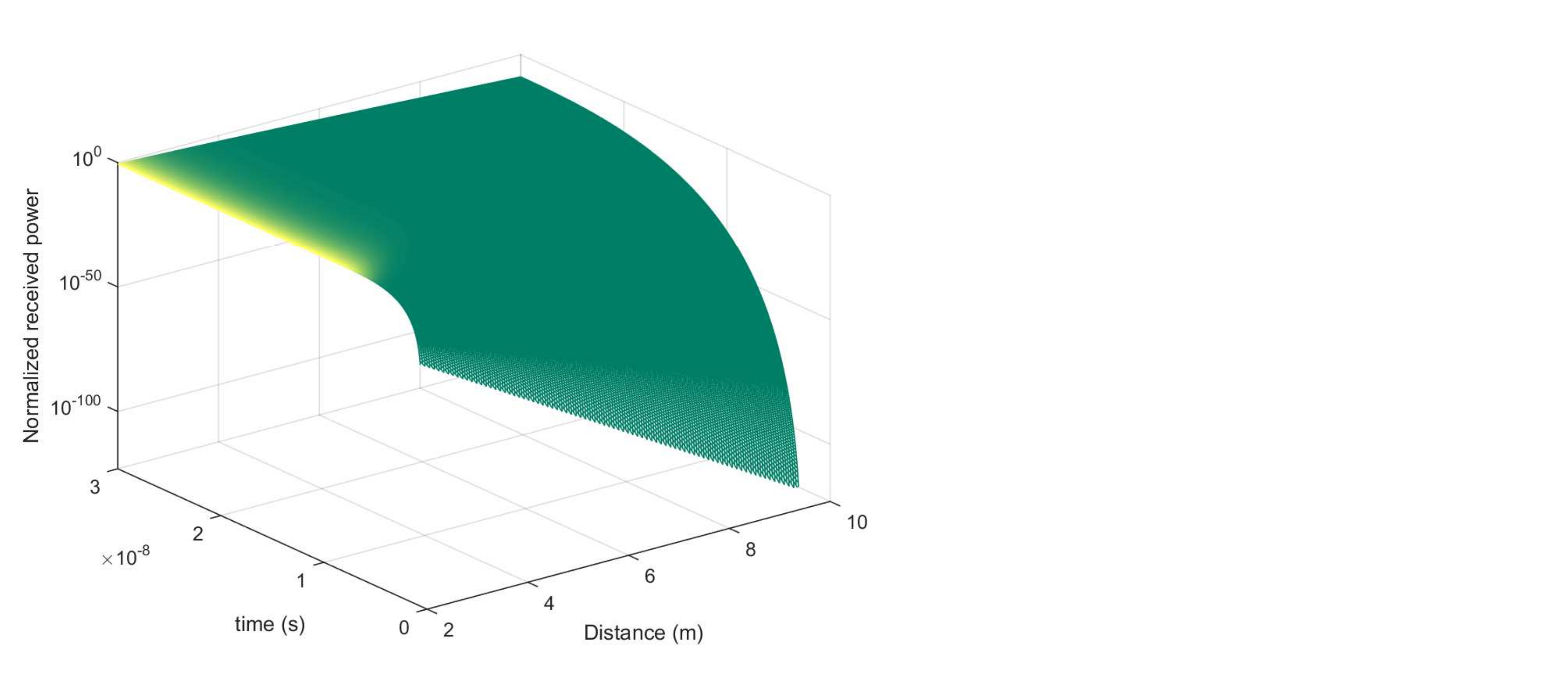}
\end{center}
\caption{Normalized received power in 3D versus distance and time,
considering TTHG\ function and a receiver aperture of 10cm.}
\label{fig12}
\end{figure}
\begin{figure}[h]
\begin{center}
\hspace*{-0.5cm}\includegraphics[scale=0.47]{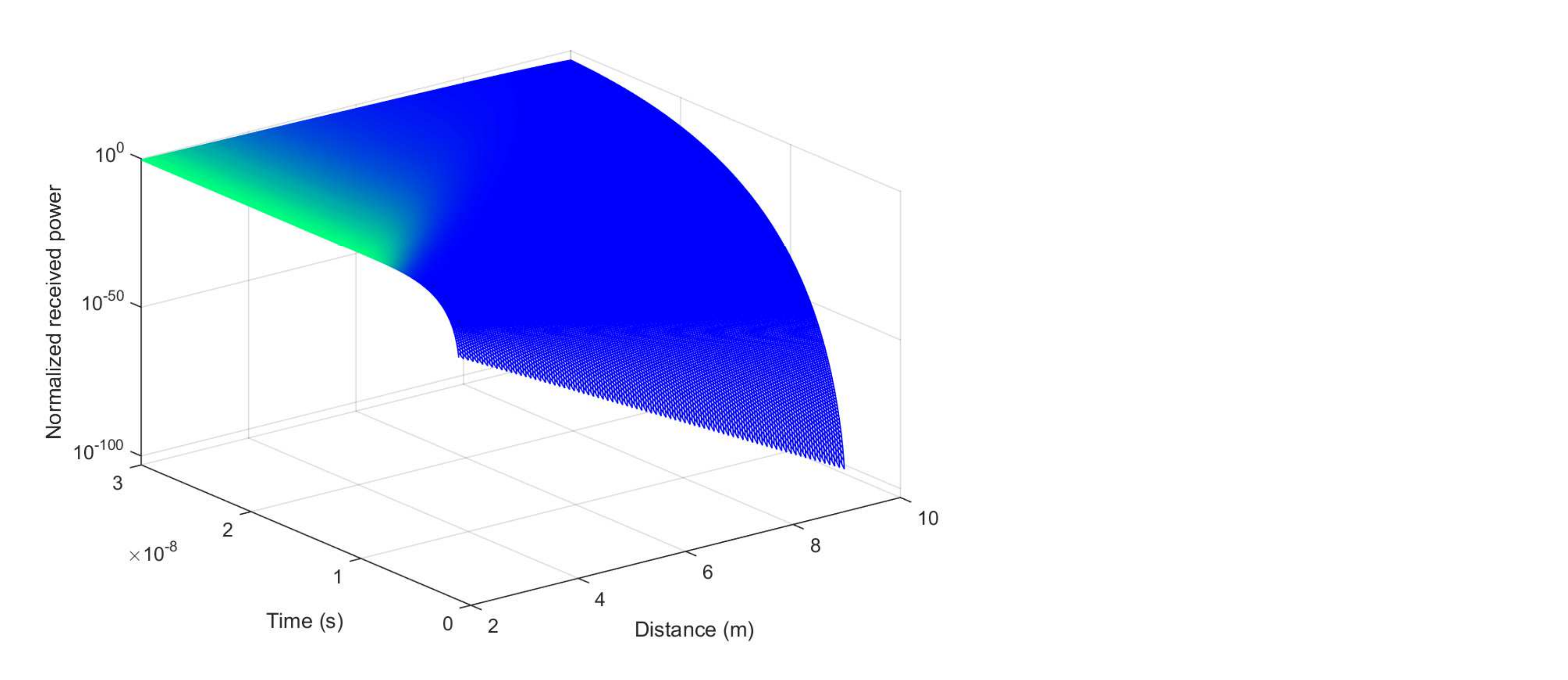}
\end{center}
\caption{Normalized received power in 3D versus distance and time,
considering FF function and a receiver aperture of 10cm.}
\label{fig13}
\end{figure}

Figs. 14-16 depict the TI-RTE\ normalized received power results in two
dimensions computed by RTE\ numerical solver as well as the respective Monte
Carlo results by considering the adopted volume scattering functions (i.e.,
STHG, TTHG, and FF). The numerical time-independent RTE\ results correspond
to the average normalized received power over time$.$ We can notice
obviously that the power loss in Harbor-I water type is greater compared to
the Harbor-II case. That is, the higher the scattering coefficient is (i.e.,
$b=0.91$ for Harbor-I and $b=1.8177$ for Harbor-II), the greater is the
path-loss, and consequently, the system performance degrades. Additionally,
for Fig. 14, one can note clearly the matching between the proposed
numerical scaled RTE curves as well as MC results considering STHG function,
on the various propagation distances values, while the RTE\ solver proposed
in \cite{alouini} yields a significant gap with MC\ curves, which proves the
computation accuracy of the proposed numerical solver. A similar behavior
can be noticed in Fig. 15 and Fig. 16 for the TTHG\ and FF scattering
functions, and more particularly for the Harbor-I water type, while a slight
gap is noticed between the proposed RTE\ solver and MC\ simulation curves at
higher distances for the Harbor-II case.

\begin{figure}[h]
\begin{center}
\hspace*{-.3cm} \includegraphics[scale=0.47]{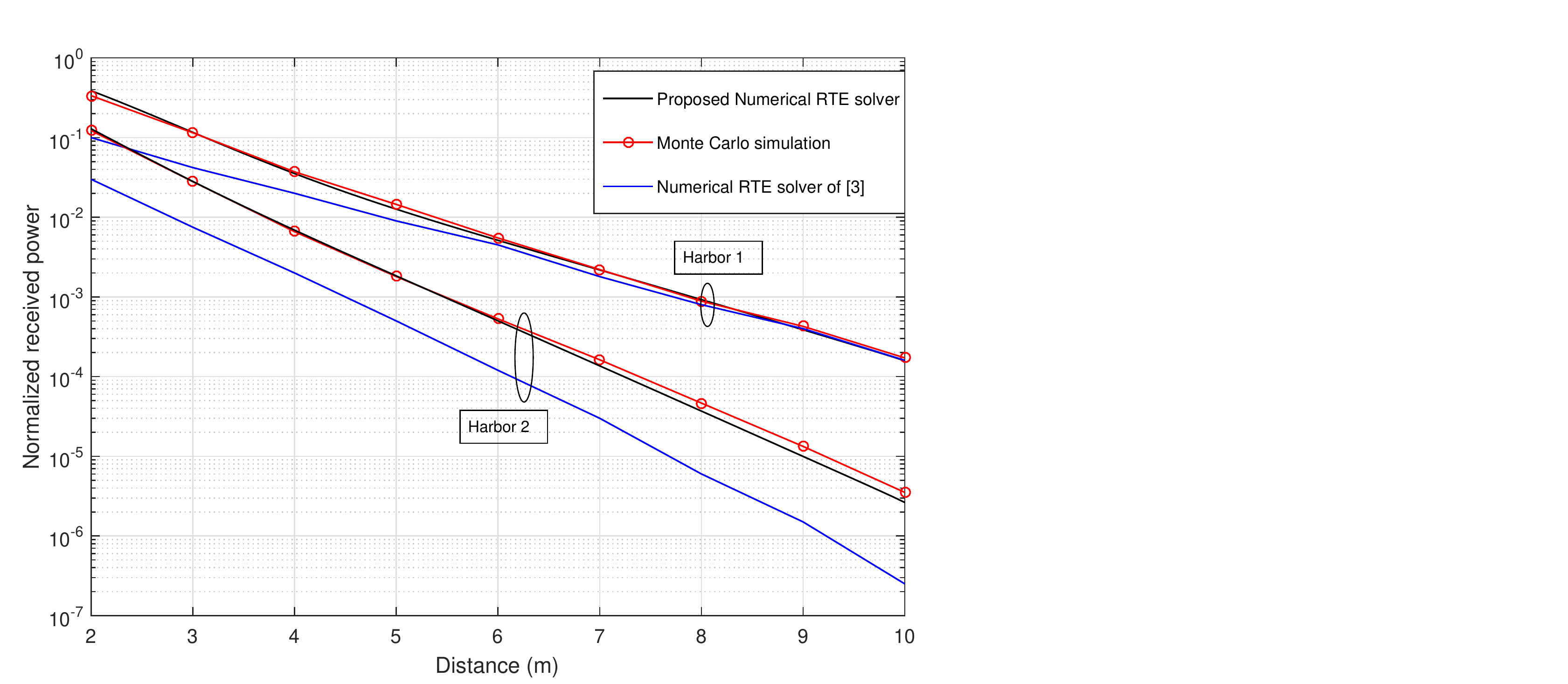}
\end{center}
\caption{Average normalized received power over time versus distance
considering STHG\ function and a receiver aperture 10cm.}
\label{fig14}
\end{figure}
\begin{figure}[h]
\begin{center}
\hspace*{-.8cm} \includegraphics[scale=0.47]{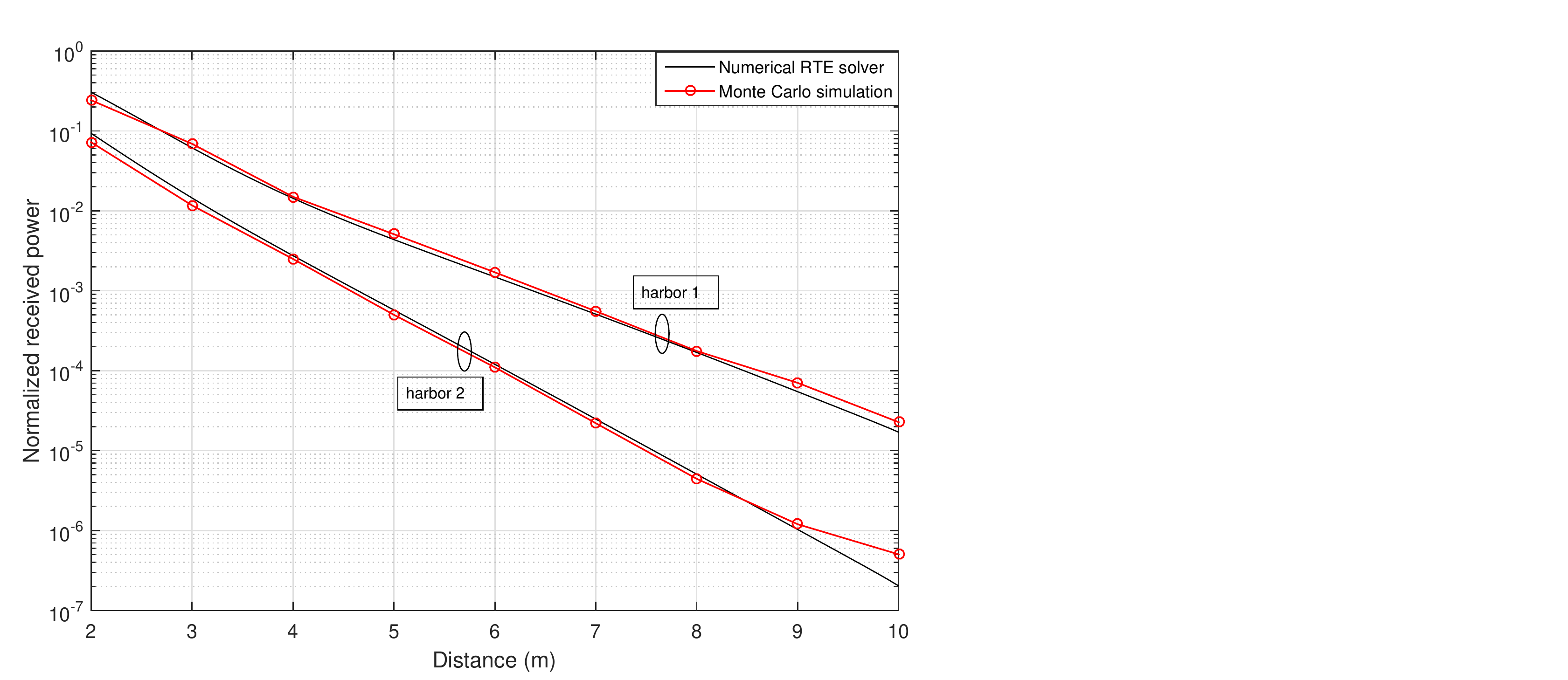}
\end{center}
\caption{Average normalized received power over time versus distance
considering TTHG function and a receiver aperture 10cm.}
\label{fig15}
\end{figure}
\begin{figure}[h]
\begin{center}
\hspace*{-1cm}\includegraphics[scale=0.47]{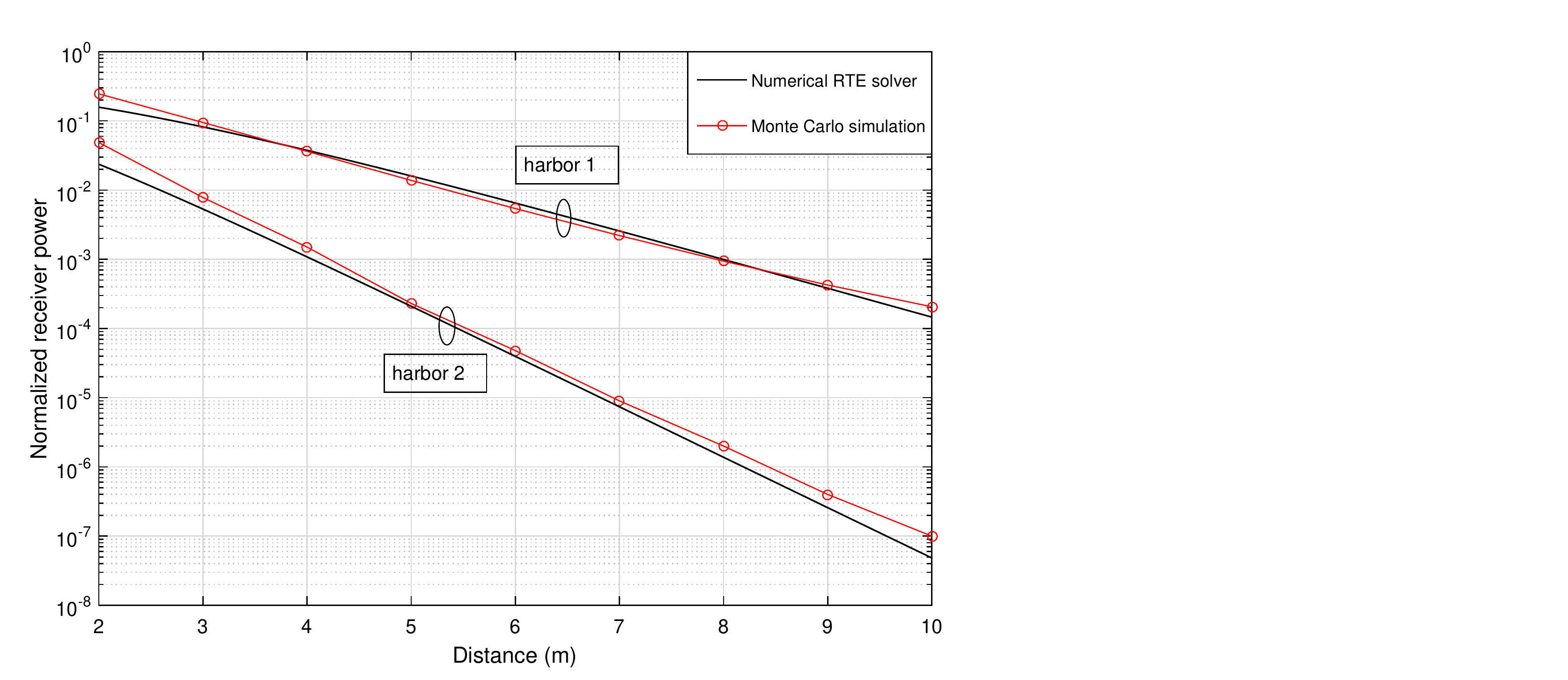}
\end{center}
\caption{Average normalized received power over time versus distance
considering FF function and a receiver aperture 10cm.}
\label{fig16}
\end{figure}

Fig. 17 depicts the total time consumption comparison between the proposed
numerical RTE\ solver and Monte Carlo simulation. One can notice clearly the
logarithmic increase of the MC\ scheme time consumption in the log-scale,
which corresponds to a linear increase in the linear scale. On the other
hand, we remark also that the computation time of the proposed RTE\ solver
as well as the one proposed in \cite{alouini} remains constant. In addition
to this, we can notice that the respective complexities of the proposed RTE
solver and the one in \cite{alouini} are very close, in terms of computation
complexity. Importantly, the main outcome of this result is the difference
in complexity between the proposed numerical RTE\ solver and its MC\
counterpart. That is, the numerical RTE\ solver can achieve accurate
results, with a remarkably reduced complexity compared to MC\ method.
\begin{figure}[h]
\begin{center}
\hspace*{-.3cm}\includegraphics[scale=0.47]{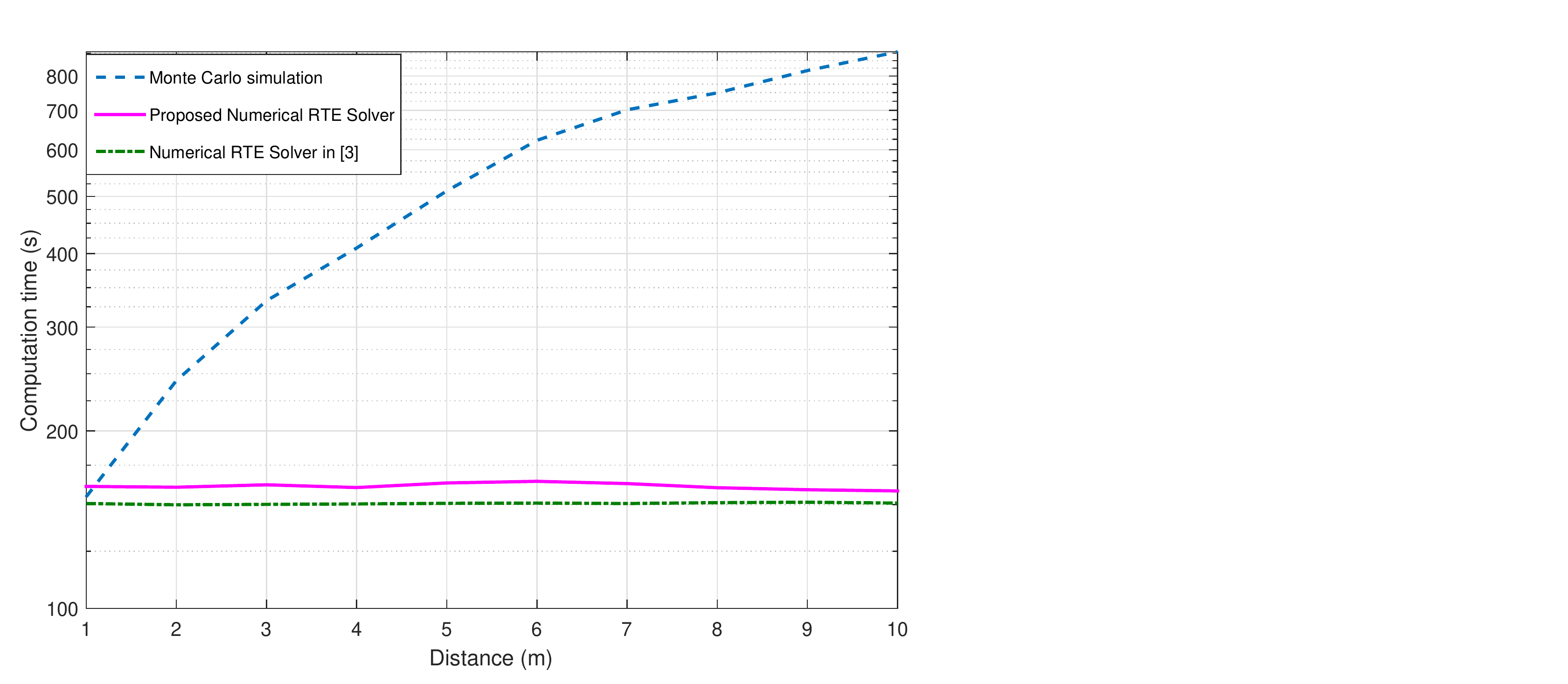}
\end{center}
\caption{Computation time consumption comparison between Numerical RTE\
Solver and Monte Carlo simulation.}
\label{fig17}
\end{figure}

Figs. 18-23 depicts the BER performance of the communication system in 2D
and 3D, based on (\ref{ber}) for both water types, for the considered VSF\
functions. The 2D curves correspond to the average normalized received power
over time$.$ In a similar manner to the power behavior, one can remark
obviously that the BER\ increases as a function of the propagation distance.
That is, the greater the propagation distance is, the more important is the
path-loss, and consequently, the BER\ performance degrades. Additionally,
the BER\ decreases with the increase of the received power as a function of
the time. Also, Harbor-I water type exhibits a lower bit error rate to
Harbor-II one.

\begin{figure}[h]
\begin{center}
\includegraphics[scale=0.47]{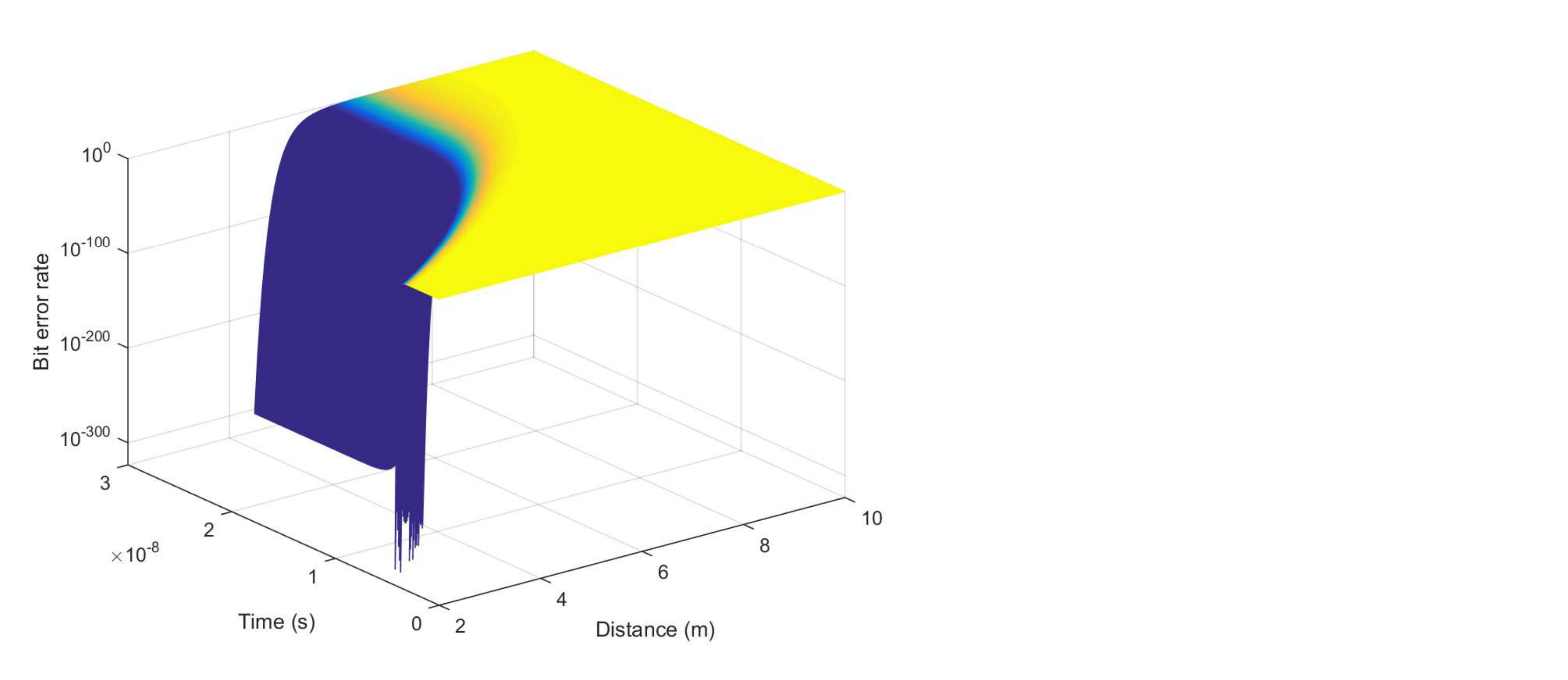}
\end{center}
\caption{Bit error rate performance of the considered system in 3D, based on
time-dependent RTE resolution considering STHG\ function (aperture: 10cm).}
\label{fig18}
\end{figure}
\begin{figure}[h]
\begin{center}
\hspace*{-.5cm}\includegraphics[scale=0.47]{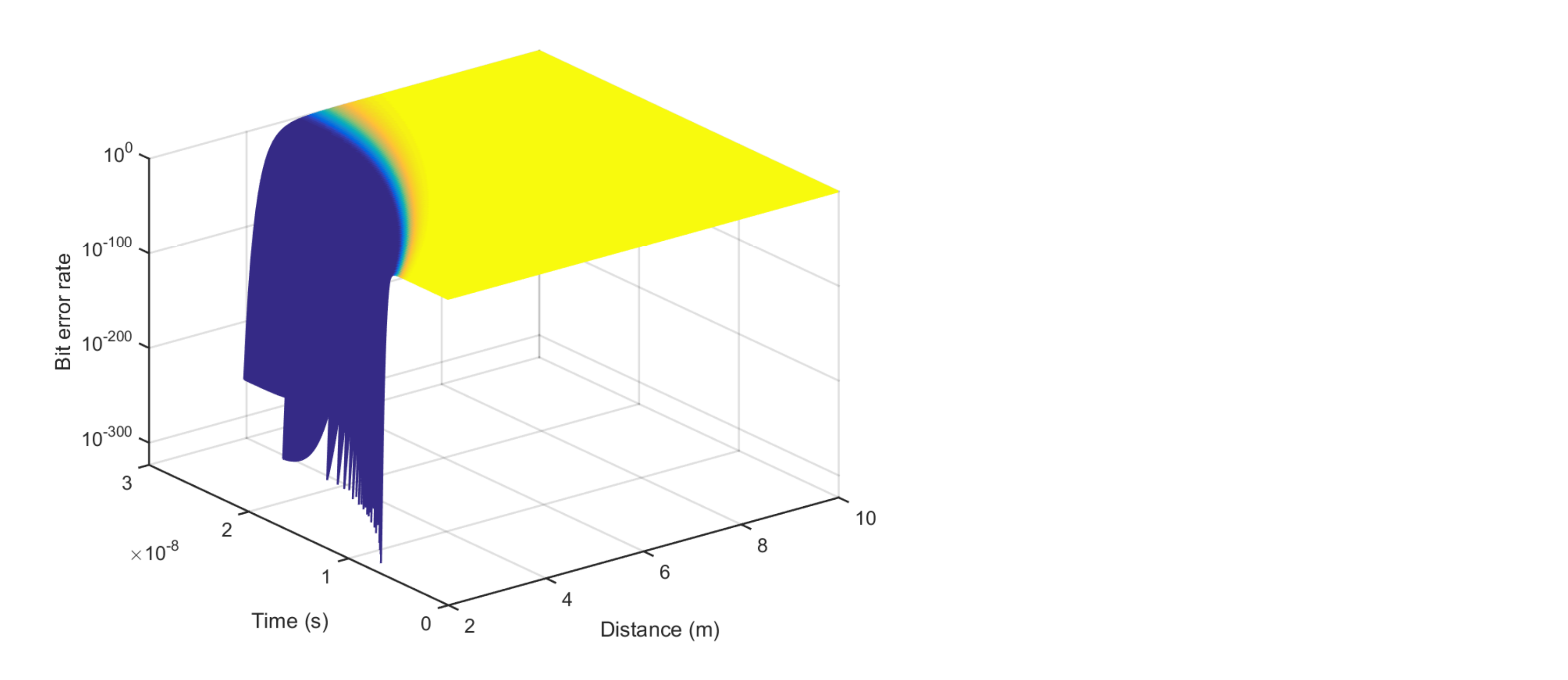}
\end{center}
\caption{Bit error rate performance of the considered system in 3D, based on
time-dependent RTE resolution considering TTHG\ function (aperture: 10cm).}
\label{fig19}
\end{figure}
\begin{figure}[h]
\begin{center}
\hspace*{-.5cm}\includegraphics[scale=0.47]{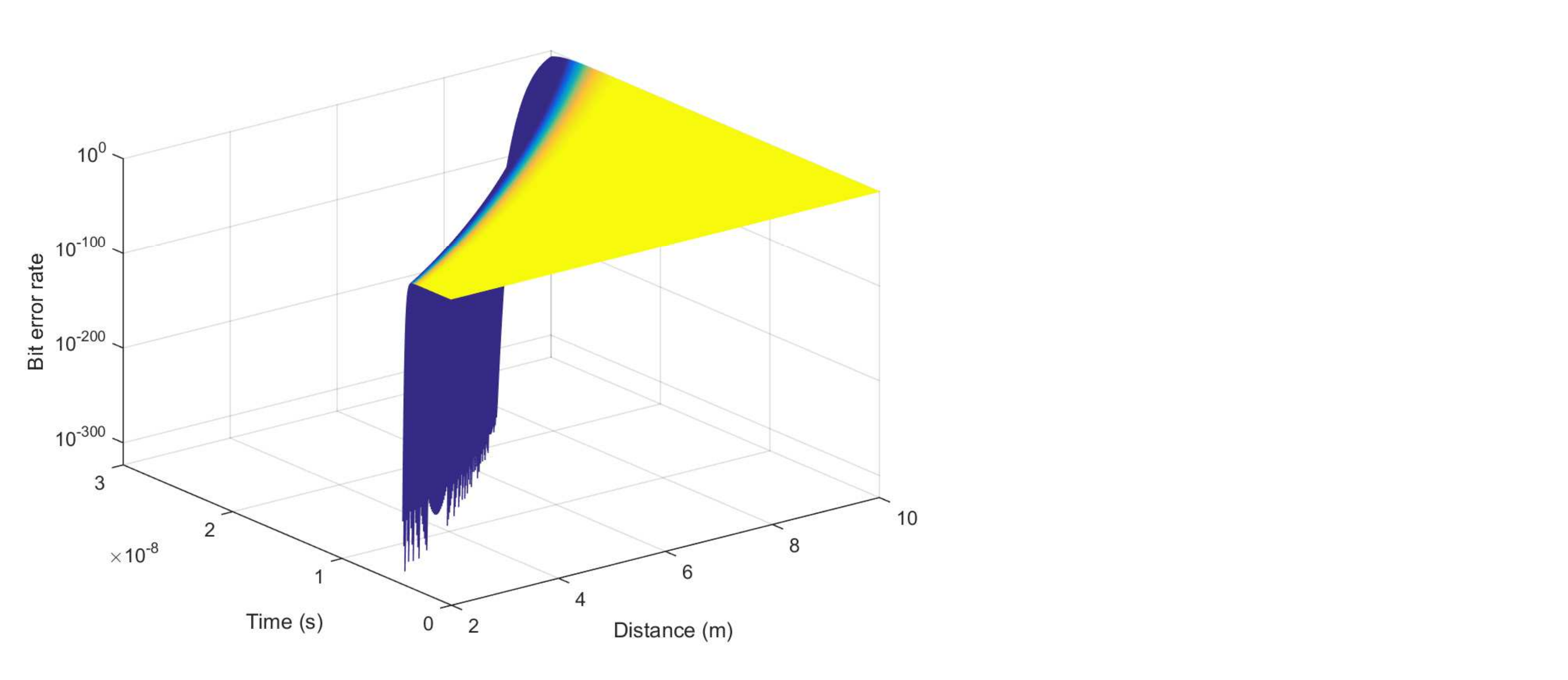}
\end{center}
\caption{Bit error rate performance of the considered system in 3D, based on
time-dependent RTE resolution considering FF\ function (aperture: 10cm).}
\label{fig20}
\end{figure}

\begin{figure}[h]
\begin{center}
\hspace*{-.5cm}\includegraphics[scale=0.47]{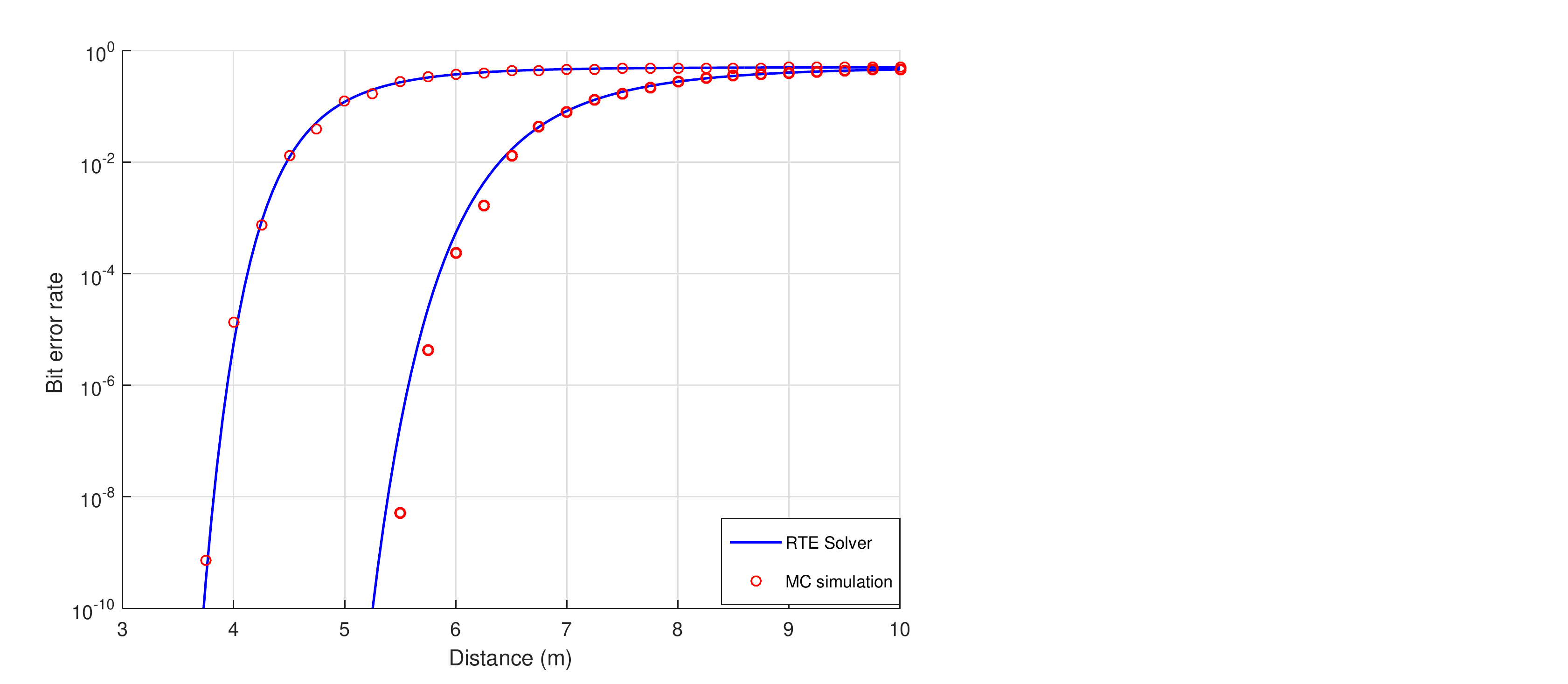}
\end{center}
\caption{Bit error rate performance of the considered system based on the
average received power over time, considering STHG\ function (aperture:
10cm). }
\label{fig21}
\end{figure}
\begin{figure}[h]
\begin{center}
\includegraphics[scale=0.47]{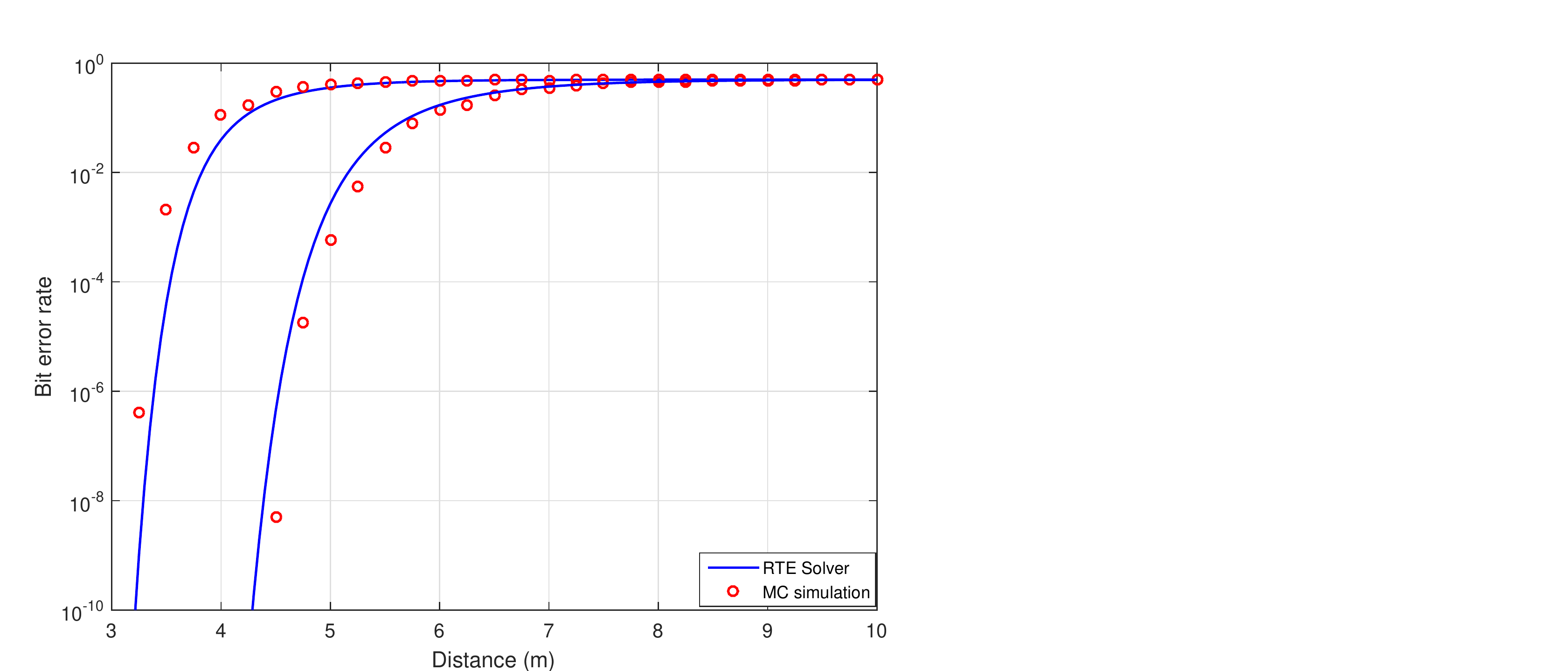}
\end{center}
\caption{Bit error rate performance of the considered system based on the
average received power over time, considering TTHG function (aperture:
10cm). }
\label{fig22}
\end{figure}

\begin{figure}[h]
\begin{center}
\hspace*{-1cm}\includegraphics[scale=0.47]{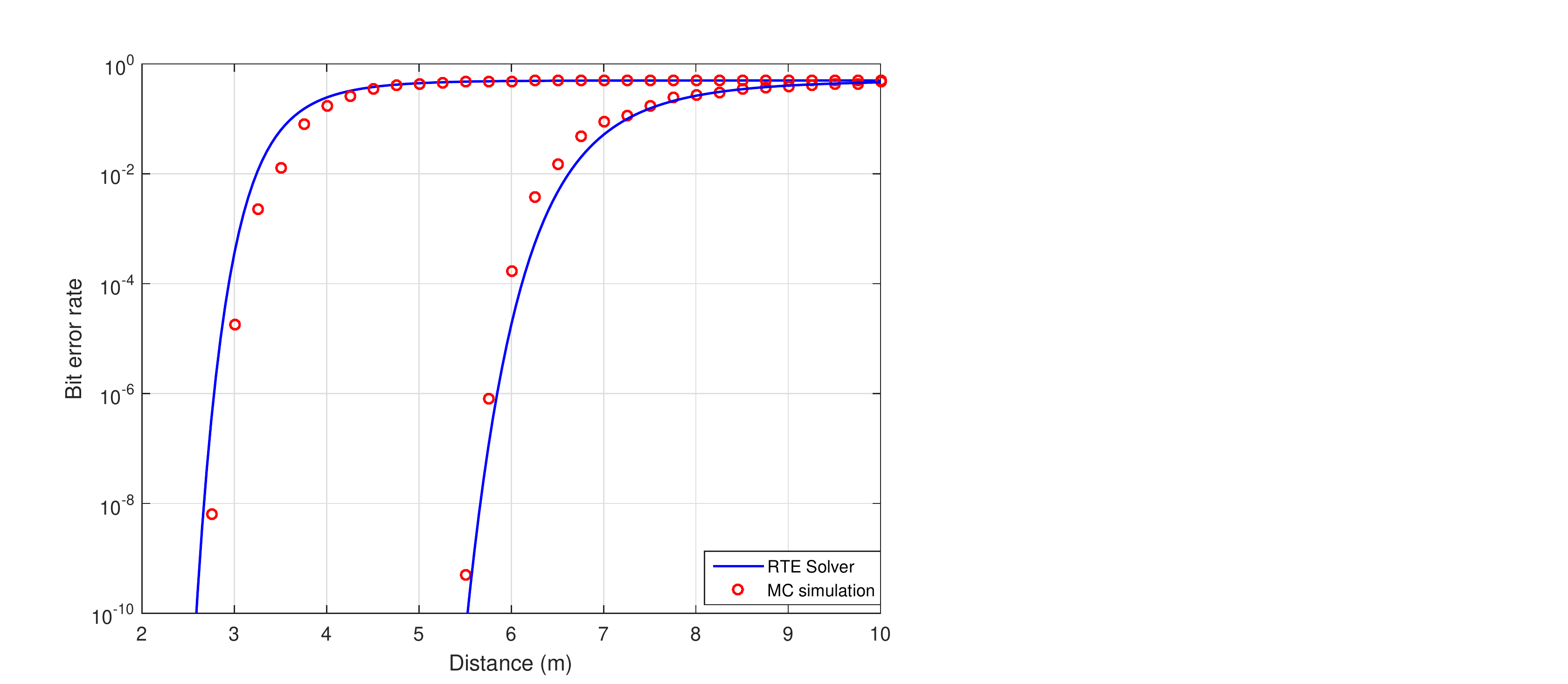}
\end{center}
\caption{Bit error rate performance of the considered system based on the
average received power over time, considering FF function (aperture: 10cm). }
\label{fig23}
\end{figure}

\section{Conclusion}

\bigskip In this paper, an improved time-dependent RTE numerical evaluation
algorithm is proposed, in order to solve the radiative transfer equation
that quantifies the light propagation loss in the underwater medium. The
proposed RTE\ solver was applied for three types of volume scattering
functions namely, the STHG, TTHG, and FF PSFs. Boole's rule given by the
5-points and 7-points Newton-Cotes formula was incorporated as a quadrature
method alongside with the two and three points Simpson's method in order to
solve the integral term. The upwind finite difference scheme was also
improved by adding one more neighbor point. Furthermore, the MSE-based
algorithm for scattering angles discretization was modified from \cite%
{alouini}, \cite{commnet}. The received light power was calculated in terms
of system and channel parameters, such as propagation distance, time,
absorption and scattering coefficient, as well as the number of angles.
Based on this result, the BER\ performance of the considered UOWC\ system
was analyzed in terms of the system and channel parameters. The proposed RTE
numerical solver and Monte Carlo simulations have been compared in terms of
tightness and complexity, where the results present a good agreement between
RTE\ and MC\ results. Furthermore, the results show that the proposed solver
has significantly less complexity compared to its MC\ counterpart. Matlab
codes of the proposed RTE solver have been presented. A potential extension
of this is considering the turbulence effects due to the dynamic change of
the pressure and temperature in the marine medium, as well as taking into
account the pointing error impairment.

\section*{Appendix: MATLAB Code for the Proposed RTE Solver}


\begin{lstlisting}
%% Main code for the RTE Solver
function y=RTE_solver(c,albedo,start,length_y,length_z,step_y,step_z,aperture,K,g,alph,g1,g2,mu,nn,ord,psf)
b=c*albedo;
I=(length_y/step_y)+1;
M=floor(aperture/step_y)+1;
r(1)=step_y/2;
s=zeros(1,(M-1)/2+1);
s(1,1)=pi*r(1)^2;
for n=2:(M-1)/2+1
r(n)=r(n-1)+step_y;
s(n)=pi*r(n)^2-s(n-1);
end
    J=(length_z/step_z)+1;
    q=zeros(I,J,K);
q(((I-1)/2)+1,1,1)=1; %% src at the middle
%% Non-Uniform discretization
phi=rte_angle_dist_org_pr_biss(g,alph,g1,g2,mu,nn,1+K/2,psf);
%% Computing weight coefficients of the integral term
 [tt,w, theta]=weight_biss_pr(phi,K,g,alph,g1,g2,mu,nn,ord,psf);
 %% iterative computing of the received radiance
 [radiance]=gauss_rte_biss_p(w,theta,length_y,length_z,step_y,step_z,K,q,b,c);
 intensity=zeros(I,J);
  phi(K+1)=2*pi+0.0015/2;
    for k=1:K
%% Field of view test
if ((phi(k)>=0 && phi(k)<=pi))
intensity(:,:)=intensity(:,:)+radiance(:,:,k)*(phi(k+1)-phi(k));
end
    end
%% power computation
for jj=1:J
powerr(jj)=s*intensity((I-1)/2+1:(M-1)/2+(I-1)/2+1,jj);
end
for ii=1:1:J-start/step_z
powertrc(ii)=powerr(ii+start/step_z);
end
 tm=toc
z=start:step_z:length_z;
semilogy(z,powertrc_7./(pi*10^-6),'r-.')
grid on;
__________________________________________________
%% Non-uniform angles discretization
function y=rte_angle_dist_org_pr_biss(g,alph,g1,g2,mu,nn,N,psf)
%% 1-initializing directions uniformly
maxit=550; % max number of iterations: MSE convergence criterion
if (strcmp(psf,'STHG')==1)
%%% STHG fct
% 2D
bta1=@(teta) teta.*(1-g.^2)./(2.*pi.*(1+g.^2-2.*g.*cos(teta)));

bta2=@(teta) (1-g.^2)./(2.*pi.*(1+g.^2-2.*g.*cos(teta)));
elseif (strcmp(psf,'TTHG')==1)
%% 3D
% %% TTHG fct
bta1=@(teta) teta.*(alph*(1-g1.^2)./(2.*pi.*(1+g1.^2-2.*g1.*cos(teta)))+(1-alph)*(1-g2.^2)./(2.*pi.*(1+g2.^2-2.*g2.*cos(teta))));

bta2=@(teta) alph*(1-g1.^2)./(2.*pi.*(1+g1.^2-2.*g1.*cos(teta)))+(1-alph)*(1-g2.^2)./(2.*pi.*(1+g2.^2-2.*g2.*cos(teta)));
%   %% Fournier-Forand fct
elseif (strcmp(psf,'FF')==1)
 v=(3-mu)/2;
dlt=@(teta) 4/(3*(nn-1)^2).*(sin(teta/2)).^2;
dlt_pi=4/(3*(nn-1)^2);
 bta1=@(teta) teta.* 1./(4*pi*(1-dlt(teta)).^2.*dlt(teta).^v).*(v.*(1-dlt(teta))-(1-dlt(teta).^v)+(dlt(teta).*(1-dlt(teta).^v)-v.*(1-dlt(teta))).*(sin(teta./2)).^(-2))+...
     (1-dlt_pi^v)./(16*pi*(dlt_pi-1).*dlt_pi^v).*(3.*(cos(teta)).^2-1);

 bta2=@(teta)  1./(4*pi*(1-dlt(teta)).^2.*dlt(teta).^v).*(v.*(1-dlt(teta))-(1-dlt(teta).^v)+(dlt(teta).*(1-dlt(teta).^v)-v.*(1-dlt(teta))).*(sin(teta./2)).^(-2))+...
     (1-dlt_pi^v)./(16*pi*(dlt_pi-1).*dlt_pi^v).*(3.*(cos(teta)).^2-1);
else
    fprintf('\n Invalid entry for the PSF\n');
end
%% 1-  Uniform distribution of angles
phi=rte_unif_dist_org(N);

%% 2- computing t(k) vector for k=1:N-1
t=zeros(1,2*(N-1)+1);
t(1)=0.0015/2; % t(0)=0


for kk=1:maxit

    for jj=1:2*(N-1)-1
   t(jj+1)=(phi(jj)+phi(jj+1))/2;
    end
    t(2*(N-1)+1)=(phi(2*(N-1))+2*pi)/2;

    for ii=1:2*(N-1)
   phi(ii)=integral (bta1,t(ii),t(ii+1))./integral(bta2,t(ii),t(ii+1));
    end

end
y=phi;
end
__________________________________________________
 %% computes the weight coefficients associated to the quadrature method
 function [tt,w,theta]=weight_biss_pr(x,N_angle,g,alph,g1,g2,mu,nn,ord,psf)
tic
w=zeros(N_angle,N_angle);
m=(N_angle+2)/2;
%%% STHG fct
if(strcmp(psf,'STHG'))
f=@(x) (1-g^2)./(2*pi*(1+g^2-2*g*cos(x))); %2D
elseif(strcmp(psf,'TTHG'))
%%% TTHG fct
 f=@(x) alph*(1-g1^2)./(2*pi*(1+g1^2-2*g1*cos(x)))+(1-alph)*(1-g2^2)./(2*pi*(1+g2^2-2*g2*cos(x)));
elseif(strcmp(psf,'FF'))
%% Fournier For. fct
v=(3-mu)/2;
dlt=@(x) 4/(3*(nn-1)^2).*(sin(x/2)).^2;
dlt_pi=4/(3*(nn-1)^2);
f=@(x)   1./(4*pi*(1-dlt(x)).^2.*dlt(x).^v).*(v.*(1-dlt(x))-(1-dlt(x).^v)+(dlt(x).*(1-dlt(x).^v)-v.*(1-dlt(x))).*(sin(x./2)).^(-2))+...
     (1-dlt_pi^v)./(16*pi*(dlt_pi-1).*dlt_pi^v).*(3.*(cos(x)).^2-1);
else
fprintf('\nInvalid entry for the PSF\n');
end

for ll=1:m-1
% for ll=1:m
h1=x(ll);
h2=x(ll+1);
M=7;
h=(h2-h1)/(M-1);
    ss=zeros(1,M);
    tt=zeros(1,M);
    uu=zeros(1,M);
    if (ord==3)
%%%%%%%%%% 3 pts scheme
    %%2 pts
    uu(1)=1/6*(2*f(h1)+f(h1+h))*h;
uu(M)=1/6*(f((M-2)*h)+2*f((M-1)*h))*h;
%% 3 points

% 3 points
for ii=2:M-1
    uu(ii)=1/12*(f(h1-(ii-2)*h)+4*f(h1+(ii-1)*h)+f(h1+ii*h))*2*h;
end
w(1,ll)=sum(uu);


    elseif (ord==5)
 %%%%% 5 pts scheme
 %%2 pts
tt(1)=1/12*(2*f(h1)+f(h1+h))*h;
tt(M)=1/12*(f((M-2)*h)+2*f((M-1)*h))*h;
%% 3 points
tt(2)=1/12*(f(h1)+4*f(h1+h)+f(h1+2*h))*2*h;
tt(M-1)=1/12*(f(h2-2*h)+4*f(h2-h)+f(h2))*2*h;

%% 5 points
for ii=3:M-2
    tt(ii)=1/360*(7*f(h1+(ii-3)*h)+32*f(h1+(ii-2)*h)+12*f(h1+(ii-1)*h)+32*f(h1+(ii)*h)+7*f(h1+(ii+1)*h))*(4*h);
end
w(1,ll)=sum(tt);

    elseif (ord==7)
% %%%%%% 7 pts scheme

 %%2 points
ss(1)=1/18*(2*f(h1)+f(h1+h))*h;
ss(M)=1/18*(f((M-2)*h)+2*f((M-1)*h))*h;
%% 3 points
ss(2)=1/36*(f(h1)+4*f(h1+h)+f(h1+2*h))*2*h;
ss(M-1)=1/36*(f(h2-2*h)+4*f(h2-h)+f(h2))*2*h;
%% 5 points
ss(3)=1/90*1/2*(7*f(h1)+32*f(h1+h)+12*f(h1+2*h)+32*f(h1+3*h)+7*f(h1+4*h))*(4*h);
ss(M-2)=1/90*1/2*(7*f(h2-4*h)+32*f(h2-3*h)+12*f(h2-2*h)+32*f(h2-h)+7*f(h2))*(4*h);
%% 7 points
for ii=4:M-3
    ss(ii)=1/840*1/6*(41*f(h1+(ii-4)*h)+216*f(h1+(ii-3)*h)+27*f(h1+(ii-2)*h)+272*f(h1+(ii-1)*h)+27*f(h1+(ii)*h)+216*f(h1+(ii+1)*h)+41*f(h1+(ii+2)*h))*(6*h);
end

w(1,ll)=sum(ss(1:M-1));

    else
        fprintf('the provided order number is neither 3, nor 5/7 !!! provide one of those integers')
    end
w(1,2*m-1-ll)=w(1,ll);

end

w(1,:)=w(1,:)/sum(w(1,:));
for ii=2:N_angle
for j=1:N_angle
w(ii,j)=w(1,abs(ii-j)+1);
end
end

theta=zeros(N_angle,2);

for ii=1:N_angle

theta(ii,1)=cos(x(ii));
 theta(ii,2)=sin(x(ii));
end
tt=toc
end
__________________________________________________
%% iterative computing of the received radiance
function [y]=gauss_rte_biss_p(w,theta,length_y,length_z,step_y,step_z,K,q,b,c)
I=(length_y/step_y)+1;
J=(length_z/step_z)+1;
radiance=zeros(I,J,K);
radiance_temp=zeros(I,J,K);

for l=1:320
for k=1:K
denominator1(k)=2*theta(k,2)/(3*step_y)+2*theta(k,1)/(3*step_z)+c;
denominator2(k)=2*theta(k,2)/(3*step_y)-2*theta(k,1)/(3*step_z)+c;
denominator3(k)=-2*theta(k,2)/(3*step_y)-2*theta(k,1)/(3*step_z)+c;
denominator4(k)=-2*theta(k,2)/(3*step_y)+2*theta(k,1)/(3*step_z)+c;

if theta(k,1)>0 && theta(k,2)>0

%%case 1
for i=1:I
for j=1:J
for n=1:K
sum1(n)=radiance(i,j,n)* w(k,n);
end

if (i==1)
    deriv_i=0;
elseif (i==2)
      deriv_i=radiance(i-1,j,k);
else
     deriv_i=radiance(i-1,j,k)+radiance(i-2,j,k);
end

if (j==1)
    deriv_j=0;
elseif (j==2)
      deriv_j=radiance(i,j-1,k);
else
     deriv_j=radiance(i,j-1,k)+radiance(i,j-2,k);
end

numerator(i,j,k)=sum(sum1)*b+(deriv_i)*(theta(k,2)/(3*step_y))+(deriv_j)*(theta(k,1)/(3*step_z))+c*q(i,j,k);
radiance_temp(i,j,k)=numerator(i,j,k)/denominator1(k);
end
end

elseif theta(k,1)<0 && theta(k,2)>0
%%case 2

for i=1:I
for j=1:J

for n=1:K
sum1(n)=radiance(i,j,n)* w(k,n);
end

if (i==1)
    deriv_i=0;
elseif (i==2)
      deriv_i=radiance(i-1,j,k);
else
     deriv_i=radiance(i-1,j,k)+radiance(i-2,j,k);
end

if (j==J)
    deriv_j=0;
elseif (j==J-1)
      deriv_j=radiance(i,j+1,k);
else
     deriv_j=radiance(i,j+1,k)+radiance(i,j+2,k);
end


numerator(i,j,k)=sum(sum1)*b+(deriv_i)*(theta(k,2)/(3*step_y))-(deriv_j)*(theta(k,1)/(3*step_z))+c*q(i,j,k);

radiance_temp(i,j,k)=numerator(i,j,k)/denominator2(k);
end
end
elseif theta(k,1)<0 && theta(k,2)<0
%%case 3
for i=1:I
for j=1:J
for n=1:K
sum1(n)=radiance(i,j,n)* w(k,n);
end

if (i==I)
    deriv_i=0;
elseif (i==I-1)
      deriv_i=radiance(i+1,j,k);
else
     deriv_i=radiance(i+1,j,k)+radiance(i+2,j,k);
end

if (j==J)
    deriv_j=0;
elseif (j==J-1)
      deriv_j=radiance(i,j+1,k);
else
     deriv_j=radiance(i,j+1,k)+radiance(i,j+2,k);
end


numerator(i,j,k)=sum(sum1)*b-(deriv_i)*(theta(k,2)/(3*step_y))-(deriv_j)*(theta(k,1)/(3*step_z))+c*q(i,j,k);

radiance_temp(i,j,k)=numerator(i,j,k)/denominator3(k);
end
end
else

%%case 4

for i=1:I
for j=1:J
for n=1:K
sum1(n)=radiance(i,j,n)* w(k,n);
end

if (i==I)
    deriv_i=0;
elseif (i==I-1)
      deriv_i=radiance(i+1,j,k);
else
     deriv_i=radiance(i+1,j,k)+radiance(i+2,j,k);
end

if (j==1)
    deriv_j=0;
elseif (j==2)
      deriv_j=radiance(i,j-1,k);
else
     deriv_j=radiance(i,j-1,k)+radiance(i,j-2,k);
end

numerator(i,j,k)=sum(sum1)*b-(deriv_i)*(theta(k,2)/(3*step_y))+(deriv_j)*(theta(k,1)/(3*step_z))+c*q(i,j,k);

radiance_temp(i,j,k)=numerator(i,j,k)/denominator4(k);
end
end
end
end
radiance=radiance_temp;
end

y=radiance;

    \end{lstlisting}

\EOD

\end{document}